\documentclass[twocolumn,prd,superscriptaddress,preprintnumbers,nofootinbib]{revtex4}[11pt]
\usepackage[usenames,dvipsnames]{xcolor}
\usepackage{amsmath, amssymb, graphicx, epsf, verbatim, hyperref, comment, color, slashed, subfigure, comment, mdwlist, paralist, rotating, booktabs, bm, multirow, topcapt}

\newcommand{\beq}{\begin{equation}}
\newcommand{\eeq}{\end{equation}}
\newcommand{\bea}{\begin{eqnarray}}
\newcommand{\eea}{\end{eqnarray}}
\newcommand{\bit}{\begin{itemize}}
\newcommand{\eit}{\end{itemize}}
\newcommand{\ben}{\begin{enumerate}}
\newcommand{\een}{\end{enumerate}}
\renewcommand{\t}{\hat}

\usepackage{aas_macros}
\usepackage{float}
\restylefloat{table}
\usepackage{soul}

\renewcommand{\eqref}[1]{Eq.~(\ref{eq:#1})}
\newcommand{\eqsref}[2]{Eqs.~(\ref{eq:#1}) and (\ref{eq:#2})}

\newcommand{\secref}[1]{Section~\ref{sec:#1}}

\newcommand{\appref}[1]{Appendix~\ref{sec:#1}}
\newcommand{\figref}[1]{Fig.~\ref{fig:#1}}

\newcommand{\tabref}[1]{Table~\ref{tab:#1}}

\newcommand{\f}{\frac}

\newcommand{\OO}{\mathcal{O}}

\newcommand{\vev}[1]{ \left\langle {#1} \right\rangle }

\newcommand{\msun}{\text{M}_\odot}

\newcommand{\pc}{\text{pc}}

\begin{document}

\title{Dynamical Evolutions in Globular Clusters and Dwarf Galaxies: \\ 
Conduction Fluid Simulations}

\author{Yi-Ming Zhong}
\thanks{yiming.zhong@cityu.edu.hk}
\affiliation{Department of Physics, City University of Hong Kong, Kowloon, Hong Kong SAR, China}

\author{Stuart L. Shapiro}
\thanks{slshapir@illinois.edu}
\affiliation{Departments of Physics and Astronomy, University of Illinois at Urbana-Champaign, Urbana, Illinois 61801, USA}

\begin{abstract} 
We present a new two-fluid conduction scheme to simulate the evolution of an isolated, self-gravitating, equilibrium cluster of stars and collisionless dark matter on secular (gravothermal) timescales. We integrate the equations in Lagrangian coordinates via a second-order, semi-implicit algorithm, which is unconditionally stable when the mass of the lighter species is much less than that of the heavier species. The method can be straightforwardly generalized to handle a multi-species system with a population of stars or components beyond collisionless dark matter and stars. We apply the method to simulate the dynamical evolution of stellar-dark matter systems, exploring the consequences of mass segregation and gravothermal core collapse, and assessing those effects for observed globular clusters and dwarf galaxies in the Local Volume.
\end{abstract}

\maketitle

\section{Introduction}
\label{sec:intro}

Globular clusters (GCs) and ultra-compact dwarf galaxies (UCDs) are self-gravitating, equilibrium systems with masses smaller than those of ordinary dwarf galaxies. GCs are collections of ancient Population II stars, with a total mass of $\OO(10^5-10^6\, \msun)$ and radii in the range $\OO(10-100\, \pc)$.  UCDs have comparable physical size, stellar composition, and luminosity to those of GCs~\cite{Mieske:2008si, misgeld2011families}. (See, e.g., GC and dwarf data compiled by the Local Volume Database (LVDB)~\cite{Pace:2024sys} for comparisons of their properties.)  GCs are formed through the collapse of molecular clouds or accretion~\cite{2018RSPSA.47470616F}. They appear to have little or no dark matter (DM), but the existence and role of DM during GC formation remain unclear.  UCDs are formed as the remnant stellar nuclei of tidally stripped dwarf galaxies~\cite{Bekki:2003qy,Pfeffer:2013mya} or merged stellar clusters~\cite{2002MNRAS.330..642F, 2002A&A...383..823M, 2012A&A...537A...3M}. They contain a significant amount of DM.
Given the similarities in the appearance of GCs and UCDs, it has also been hypothesized that these two types of self-gravitating systems may have originated from the same type of progenitor galaxy through dynamical evolution processes such as dynamical friction, mass segregation, tidal stripping, or a combination of these processes~\cite{2023MNRAS.522.1726V, 2024MNRAS.529.4104V, Baumgardt:2008abc, Ardi:2020hlx}.

To identify the formation mechanism of GCs and UCDs, track their evolution, and confront them with 
observational data, robust simulations are needed. Most previous studies rely on N-body simulations (e.g.,~\cite{1992MNRAS.257..513H, Giersz:1994tv, Baumgardt:2002ya, Heggie_Hut_2003,  Baumgardt:2008abc, Breen:2011zy, 2017MNRAS.472..744B, Ardi:2020hlx, 2022MNRAS.510.3531B}), which typically evolve
a large representative sample of particles interacting according to Newtonian gravity. 
Although the method is 3D and based on first principles, it suffers from the need for large computational resources to handle large N-body systems. Challenges include space and time resolution issues for close encounters, tracking secular evolution on relaxation timescales, and fluctuations. These limitations are particularly problematic in the study of GCs and UCDs because they are dense systems, especially in their central regions. To overcome those shortcomings, alternative simulation methods, including 
finite-difference and Monte-Carlo techniques based on the Fokker-Planck (FP) approximation, have been developed to
follow their secular evolution on relaxation timescales 
once they reach dynamical equilibrium (see, e.g.~\cite{Lightman:1978zz, 1987degc.book.....S} for reviews). In addition, treatments based on the conduction fluid approximation have been
used to model GCs and UCDs. As in the case of N-body simulations, conduction fluid simulations
can handle both the formative dynamical epochs and the subsequent secular phases 
of the evolution~\cite{lynden1980consequences, 1992MNRAS.257..513H, 1995MNRAS.272..772S, Balberg:2002ue, Shapiro:2018vju}.

In this paper, we present a new code based on 
the two-component conduction fluid method and apply it to simulate the dynamical evolution of collisionless DM and stars in GCs and UCDs once these clusters have settled into dynamical equilibrium. The heating leads to
mass segregation, which occurs when a heavier component (e.g., stars)  and a lighter component (e.g., sub-solar-mass primordial black holes (PBHs) or axion dark matter granules ) interact by pure gravitational 
scattering in a self-gravitating system in dynamical, but not secular, equilibrium. The heavier component drifts to the center of the system, releasing its gravitational potential energy, which heats up the lighter component and causes it to expand. The timescale on which the dynamical heating is effective and the degree of mass segregation depend on the properties of the progenitor system, such as the particle mass of each of the components, their mass ratios, and distributions.  Recently, the process has been investigated for ultra-faint dwarfs to constrain PBHs, massive compact halo objects, and dark clumps~\cite{Graham:2023unf, Graham:2024hah}, in which estimates are presented but without accounting for the evolution of the density profiles during the process. One question we ultimately hope to address is whether or not GCs, currently believed to be absent any DM, might have been formed with substantial collisionless DM that has since evaporated due to gravitational heating.

Our two-fluid conduction scheme is spherically symmetric and 
adopts distinct, finely-spaced Lagrangian zones for each of the two components 
of the system. As a result, the method overcomes the resolution issues of the N-body simulations by working with a reduced phase space and employing adaptive grids. Additionally, the conduction fluid  approach  can be easily  generalized in 
several ways, such as treating  multi-component systems, incorporating 
stellar binaries or massive central black holes, or including
external environmental effects such as tidal interactions. The approach can
straightforwardly
replace cold dark matter (CDM) with collisional or self-interacting dark matter (SIDM, \cite{Balberg:2001qg, Balberg:2002ue}, see review~\cite{Tulin:2017ara}), atomic dark matter (\cite{Kaplan:2009de, Kaplan:2011yj}, see review~\cite{Cline:2021itd})  
or dissipative dark matter~\cite{Boddy:2016bbu,Essig:2018pzq, Huo:2019yhk,Roy:2023zar}.
The ability to generalize the scheme is important since the investigation 
reported here is quite preliminary,
as it does not yet include  some potentially important effects, such as binaries and tidal interactions, that we plan to incorporate in future studies.

By applying our two-fluid code to evolve systems of GCs and UCDs, we observe that all systems display a certain degree of mass segregation. During this process, the lighter component expands outward while the heavier one collapses toward the center. As the system evolves, a dense core of the heavier component emerges in the system's center. The outer region of this core exhibits a cuspy density profile with a log-slope of approximately $-2.2$. The time it takes for significant mass segregation and core collapse to occur largely depends on the relaxation timescale of the heavier component. We examine a variety of systems, each with different ratios of total mass  and scale radius between the two components. We find that while the core collapse of the heavier component is universally manifested across all systems, the expansion of the lighter component is significant in systems whose lighter component is initially compact and represents a smaller fraction of the total mass.

By comparing with the observational data~\cite{1996AJ....112.1487H, Harris:2010ut, Pace:2024sys} in the Local Volume, we find that the majority of GCs should have experienced significant dynamical evolution, whereas most UCDs do not, apart from those with the lowest mass and highest compactness. For GCs that underwent dynamical evolution, if the DM particle mass is significantly smaller than the stellar mass, DM is heated from the central region and expands to the outer region, resulting in a stellar-dominated high-density core. Heating alone is unlikely to deplete all DM from GCs, especially at larger radii. Conversely, the opposite can occur for small UCDs housing a DM particle mass much larger than the stellar mass, giving rise to an ultra-heavy DM-dominated high-density core in those dwarfs.

The remainder of the paper is organized as follows. We sketch our simulation method in \secref{simulation} and validate it in \secref{validation}. In \secref{results}, we simulate a system with stars and sub-solar-mass or sup-solar-mass DM particles, comparing our results to N-body simulations in the literature. We also present simulations for various parameters of stellar-DM systems.
In \secref{observation}, we apply our simulation results to the observed GCs and dwarfs in the Local Volume and provide a preliminary answer to what likely happens to DM in GCs, assuming that
significant DM may have been present at the formation of such systems (which is not fully resolved by
early universe simulations). We conclude in \secref{conclusion}. \appref{CFL} provides 
further details of our algorithm for heat conduction and dynamical heating or cooling, 
showing that it always satisfies the Courant-Friedrichs-Lewy (CFL) condition for numerical
stability. \appref{trans} gives the translation of fiducial units from ours to those adopted in the literature. \appref{FP} compares the results from the fluid model and those from the FP model.

\section{Simulation Method}
\label{sec:simulation}
We consider progenitor galaxies consisting of a lighter component and a heavier component, both treated as nonrotating, spherical conducting fluids residing in a single gravitational potential in dynamical equilibrium. Such two-component systems are ubiquitous in halo modeling. For instance, the heavier component can be solar-mass stars, while the lighter component can be sub-solar-mass PBHs, a potential ultra-heavy DM candidate. Alternatively, the heavier component can be super-solar-mass PBHs, and the lighter component can be solar-mass stars.

At first glance, the fluid treatment may seem inappropriate for evolving weakly collisional, self-gravitating systems such as large $N$-body clusters with mean-free paths longer than system radii. For such systems,
the gravitational relaxation time is $\mathcal{O}(N / \ln (\gamma N))$ times the dynamical time (here $\gamma$ is a constant of order unity discussed in Sec.~\ref{coulomb} below). However, earlier studies, such as~\cite{lynden1980consequences}, 
show that the fluid method, accounting for
conduction, is in reasonable agreement with results from other numerical simulation methods~\cite{1987degc.book.....S, giersz1994statistics, spurzem1996fluid}. In particular, the fluid method has been calibrated against N-body simulations, e.g.~\cite{Koda:2011yb}, and FP integrations~\cite{1995MNRAS.272..772S}
for gravothermal collapse, with good agreement. We therefore adopt this method here. The following sections derive the two-fluid equations and then describe the numerical implementation in detail.

\subsection{Two-fluid setups}
We model both heavier and lighter components of the system as monatomic fluids with adiabatic index $\Gamma = 5/3$. Their velocity distributions follow the Maxwell-Boltzmann distribution locally. The equations of state for the heavier component (with subscript $h$) and the lighter component (with subscript $l$) are  expressed as,
\begin{align}
p_h ={}& (\Gamma-1) \rho_h u_h = \rho_h \sigma_h^2,\\
p_l ={}& (\Gamma-1) \rho_l u_l = \rho_l \sigma_l^2,
\end{align}
where $p$, $\rho$, $\sigma$, and $u \equiv {3\sigma^2}/{2}$ are the pressure, density, 1D velocity dispersion, and kinetic energy per unit mass for each fluid component.

Three types of interactions are relevant for the evolution of the two-fluid system: gravitational interactions between the heavier particles, those between the lighter particles, and those between the heavier and lighter particles. The first two give rise to conduction within each fluid, and the last one introduces heat-exchange interactions between the two. 

Following Eq.~(3-38) of~\cite{1987degc.book.....S}, the conduction equation is,
\begin{equation}
\frac{L_i}{4\pi r^2} = - \kappa_i \frac{d \sigma_i^2}{dr}, \quad \kappa_i = \beta_i \frac{\alpha b }{\sqrt{3}} \frac{G \rho_i m_i \ln \Lambda_{i}}{\sigma_i}, \quad i = l, h
\label{eq:con}
\end{equation}
where $L$ is the luminosity, $\kappa$ is the conductivity, $m$ is the particle mass, and $G$ is the gravitational constant. The constants $\alpha = 2\,\text{erf}(\sqrt{{3}/{2}}) -2\, e^{-3/2} \sqrt{{6}/{\pi }}\simeq 1.217$ and $b \simeq 0.45$. Note that Ref.~\cite{lynden1980consequences} uses a symbol $C$ for the pre-factor of $\kappa$, which is related to $\alpha$ and $b$ by  $C = \alpha b/3^{3/2} \simeq 0.105$.  $\beta_{l,h}$ are $\mathcal{O}(1)$ coefficients that have been introduced to match the fluid simulations with the N-body simulations.

For the dynamical heating process (Eq.~(2-60) of~\cite{1987degc.book.....S}, also see~\cite{Graham:2023unf}), the total kinetic energy change rate of the lighter fluid is
\begin{equation}
n_l \frac{d \langle E_l \rangle}{dt} = \frac{4 (2\pi)^{1/2} G^2 \rho_h \rho_l \ln \Lambda_{hl}}{(\sigma_l^2 + \sigma_h^2)^{3/2}} \left(m_h \sigma_h^2 - m_l \sigma_l^2 \right) \equiv R,
\label{eq:dyn}
\end{equation}
where $n$ is the number density and $\langle E \rangle$ is the amount of  specific kinetic energy change. Energy conservation requires the total kinetic energy rate-of-change of the heavier fluid to be given by $n_h \frac{d \langle E_h \rangle}{dt} = -R$.

\subsection{The Coulomb logarithms}
\label{coulomb}
The Coulomb logarithms, $\ln \Lambda_h$ and $\ln \Lambda_l$ from~\eqref{con} and $\ln \Lambda_{hl}$ from~\eqref{dyn}, result from regulating the phase space of the momentum transfer cross section for gravitational interactions, which would otherwise result in logarithmic divergences. In terms of the impact parameter $b$'s, the Coulomb logarithms, to a good approximation, are given by
\begin{equation}
\ln \Lambda \simeq \ln \frac{b_{\max}}{b_{90}} = \ln \frac{b_{\max} v_{\text{rel},ij}^2}{G(m_i +m_j)},\quad i,j = l, h
\label{eq:Coulomb}
\end{equation}
where $b_{\max}$ is the maximum impact parameter, which we  approximate it to be the half-mass radius of the halo $r_{50\%}$. $b_{90} \equiv G (m_i + m_j)/v^2_{\text{rel},ij}$ is the impact parameter for a 90$^\circ$ deflection, and $v^2_{\text{rel},ij}\equiv  (\bm v_i - \bm v_j)^2 $ is the relative 3D velocity squared between the incoming particles. Note that we set the minimum impact parameter to the value for a $90^{\circ}$ scattering ~\cite{1987degc.book.....S}, as we are focused on relaxation due to repeated, distant, small-angle encounters. A more detailed treatment of the Coulomb logarithm for different compact objects can be found in~\cite{Graham:2023unf, Graham:2024hah}. $v^2_{\text{rel}}$ can be evaluated using the thermal average of the fluid~\cite{Dvorkin:2013cea}, 
\begin{equation} 
v^2_{\text{rel},ij} \simeq \langle (\bm v_i - \bm v_j)^2 \rangle = 3 (\sigma_i^2 + \sigma_j^2).
\end{equation}
In principle, $\sigma_i$ is a function of both position and time. However, since the logarithm function is slowly varying with its argument, we replace $\sigma_i^2$ with the virial velocities  $\langle \sigma_i^2 \rangle$. The latter is given by
\begin{equation}
3 M_{\text{tot},i} \langle \sigma_i^2 \rangle = 4\pi G \int_0^\infty dr\, r \rho_i (r) \left(M_l (r) +M_h(r)\right)
\label{eq:virial}
\end{equation}
(Eq.~(1-9) of \cite{1987degc.book.....S}).
Here, $M(r)$ is the enclosed mass within radius $r$ and $M_{\text{tot}, h(l)}$ is the total mass of the heavier (lighter) component. Note that the left-hand side of~\eqref{virial} is twice the kinetic energy of all the heavier or lighter particles, and the right-hand side is the absolute value of the gravitational potential energy of  the particles. 

If we further fix the density profiles to be the Plummer profiles~\cite{Plummer:1911zza} (see~\secref{ic} below) with the same scale radius $r_s\equiv r_{sh} = r_{sl}$. We then find that
\begin{align}
r_{50\%h} {}& = r_{50\%l} \approx 1.3\, r_s, \label{eq:plummer_rs}\\
\langle \sigma_h^2 \rangle ={}& \langle \sigma_l^2 \rangle = \frac{\pi G M_\text{tot}}{32 r_s}, \label{eq:plummer_sigma}
\end{align}
where $M_\text{tot} \equiv \sum_{i=l,h} M_{\text{tot}, i}$ is the total mass of the system. Substituting~\eqref{plummer_rs} and~\eqref{plummer_sigma} into~\eqref{Coulomb}, the resulting Coulomb logarithms are given by
\begin{align}
\ln \Lambda_h \approx &{}\ln \left(\gamma \frac{ M_\text{tot}}{m_h}\right),\quad \ln \Lambda_l \approx  \ln \left(\gamma \frac{ M_\text{tot}}{m_l}\right),\nonumber \\
\ln \Lambda_{hl} \approx &{}\ln \left(\gamma\frac{ 2 M_\text{tot}}{m_l + m_h}\right),
\label{eq:log}
\end{align}
with $\gamma = 0.4$. This is the same value of $\gamma$  suggested by Ref.~\cite{1987degc.book.....S} for the one-fluid model. The value of $\gamma$ was later updated to $\gamma = 0.11$ by comparing the result of the fluid simulation to that of the N-body simulation~\cite{giersz1994statistics}. Values of $\gamma$ for other setups are reported in~\cite{Giersz:1995tma, Freitag:2005yc}.

If we relax the restriction on the scale radius and allow $r_{sh} \neq r_{sl}$, we can still numerically solve for $\vev{\sigma_h}^2$, $\vev{\sigma_l}^2$, and $r_{50\%}$ using~\eqref{virial} and the mass profile of the halo. Generally speaking, this will lead to the $\gamma$’s for $\ln \Lambda_h$, $\ln \Lambda_l$, and $\ln \Lambda_{hl}$ of~\eqref{log} to be different. For the density profile (Plummer) and the parameter space we considered ($0.5\leq r_{sl}/r_{sh}\leq 2$, $0.1\leq \rho_{sl}/\rho_{sh}\leq 10$) in~\secref{results}, the differences in $\gamma$ are up to a factor of a few times that of the scenario $r_{sh}=r_{sl}$. Such differences lead to a minor effect after taking the logarithm. Furthermore, we will mostly reside in the mass hierarchy
\beq
M_\text{tot} \gg m_h \gg m_l
\label{eq:hierarchy}
\eeq
in the simulations below. Given this hierarchy, the difference in $\gamma$ should have a minor effect on the simulation results. Therefore, we will fix $\gamma=0.11$ in the computation of the Coulomb logarithms (\eqref{log}) below. We also note that Refs.~\cite{1992MNRAS.257..513H, Graham:2023unf, Graham:2024hah} impose different choices for the argument of the Coulomb logarithms, which again will not significantly change the results because of the slow-varying property of the logarithm function. Note that~\eqref{hierarchy} implies 
\beq
\ln \Lambda_h \approx \ln \Lambda_{hl}.
\eeq

\subsection{Two-fluid equations}

Under spherical symmetry, the dynamics of the heavier and lighter fluids are governed by a set of $2\times 4$ 1D partial differential equations. These equations can be grouped into four categories: mass conservation, linear momentum conservation, energy conservation, and conduction.
\begin{enumerate}
\item \textbf{Mass Conservation:}
\begin{align}
\frac{\partial M_h}{\partial r} ={}& 4\pi r^2 \rho_h,\\
\frac{\partial M_l}{\partial r} ={}& 4\pi r^2 \rho_l;
\label{eq:1}
\end{align}
\item \textbf{Pressure Balance:}
\begin{align}
\frac{\partial}{\partial r}\left( \rho_h \sigma_h^2 \right) ={}& -\frac{G (M_l + M_h) \rho_h }{r^2},
\label{eq:21}\\
\frac{\partial}{\partial r}\left( \rho_l \sigma_l^2 \right) ={}& -\frac{G (M_l + M_h) \rho_l }{r^2};
\label{eq:22}
\end{align}
\item \textbf{Energy Conservation:}
For the heavier component,
\begin{align}
\rho_h \sigma_h^2 & \left(\frac{D}{D t}\right) \ln\frac{\sigma_h^3}{\rho_h} =  -\frac{1}{4\pi r^2}\frac{\partial L_h}{\partial r} - R
\label{eq:4_1}
\end{align}
For the lighter component,
\begin{align}
\rho_l \sigma_l^2 & \left(\frac{D}{D t}\right) \ln\frac{\sigma_l^3}{\rho_l} =  -\frac{1}{4\pi r^2}\frac{\partial L_l}{\partial r} + R
\label{eq:4_2}
\end{align}
where $D/D t$ is the Lagrangian time derivative~\cite{1992MNRAS.257..513H}. Note that the first and second terms on the right-hand sides of \eqsref{4_1}{4_2} are the gravothermal conduction and dynamical heating/cooling exchange terms, respectively;
\item \textbf{Conduction Equations:}
\begin{align}
\frac{L_h}{4\pi r^2} ={}& - \beta_h \frac{\alpha b G \rho_h m_h \ln \Lambda_h}{\sqrt{3} \sigma_h} \frac{\partial \sigma_h^2}{\partial r}, \\
\frac{L_l}{4\pi r^2} ={}& - \beta_l \frac{\alpha b G \rho_l m_l \ln \Lambda_l}{\sqrt{3} \sigma_l} \frac{\partial \sigma_l^2}{\partial r},
\label{eq:conduction}
\end{align}
\end{enumerate}
We leave a detailed calibration for future work and set 
\beq
\beta_l = \beta_h = 1
\eeq
hereafter. As we will see in~\secref{check3}, such a parameter choice already yields reasonable agreement for the evolved properties of the system between N-body simulations and the conduction fluid integrations.

Given the mass hierarchy~\eqref{hierarchy}, we can simplify~\eqref{conduction} to
\begin{equation}
L_l = 0,   {\hspace{8 mm}} (m_h \gg m_l)
\end{equation}
and~\eqref{4_2} to
\begin{equation}
\rho_l \sigma_l^2 \left(\frac{D}{D t}\right) \ln\frac{\sigma_l^3}{\rho_l} = R,  {\hspace{8 mm}}(m_h \gg m_l).
\end{equation}
The equation implies that the lighter fluid behaves as a non-conducting fluid whose evolution is entirely controlled by the dynamical heating.

\subsection{Boundary conditions}
\label{sec:bc}
We set up the following conditions for the inner boundary of the progenitor galaxy, $r=0$, 
\begin{align}
\f{\partial \rho_h}{\partial r}  ={}& 0,\quad  \f{\partial \rho_l }{\partial r} = 0,\label{eq:reg1}\\
\f{\partial \sigma_h}{\partial r}  ={}& 0,\quad  \f{\partial \sigma_l}{\partial r}  = 0,\label{eq:reg2}\\
M_h ={}& 0, \quad M_l = 0,\\
L_h ={}& 0, \quad L_l =0,
\end{align}
where the first two rows of equations ensure regular behavior near the origin during the evolution. We fix the outer boundary to a radius far beyond where any expanding species ever exhibits appreciable density (e.g., $r_{\rm outer} = 100$ times the initial Plummer scale radius of the lighter fluid). There, we set the following conditions:
\begin{eqnarray}
\rho_h (r_\text{outer}) =&{} 0 = 
\rho_l (r_\text{outer}),\\
L_{h} (r_\text{outer}) =&{} 0 = 
L_l (r_\text{outer}),
\end{eqnarray}
which ensures  the density, pressure, and luminosity vanish at the outer boundary.

\subsection{Initial conditions}
\label{sec:ic}
For the initial conditions, we assume that both the density distributions of the heavier and lighter components follow Plummer profiles,
\begin{equation}
\rho_h (r) = \frac{\rho_{sh}}{\left(1+r^2/r_{sh}^2\right)^{5/2}},\quad \rho_l (r) = \frac{\rho_{sl}}{\left(1+r^2/r_{sl}^2\right)^{5/2}},
\label{eq:plummer}
\end{equation}
where $\rho_{sh (sl)}$ and $r_{sh (sl)}$ are the scale density and scale radius for heavier (lighter) components, respectively. 
Note that our simulation formalism is applicable to other choices of initial density profiles ( e.g.,~\appref{FP}). We further define the initial scale density ratio and the scale radius ratio between the two components as
\beq
\xi \equiv \frac{\rho_{sl}}{\rho_{sh}}, \quad \zeta \equiv \frac{r_{sl}}{r_{sh}}.
\eeq
For the Plummer profile, the scaled density and radius are related to the total mass by
\beq
M_{\text{tot}, h} = \frac{4\pi}{3} \rho_{sh}r_{sh}^3, \quad M_{\text{tot}, l} = \frac{4\pi}{3} \rho_{sl}r_{sl}^3.
\label{eq:plummer_totalmass}
\eeq
Therefore, the total mass ratio of the two fluids is given by
\beq
\frac{M_{\text{tot},l}}{M_{\text{tot},h}} = \xi \zeta^3.
\eeq
The enclosed mass profiles are given by
\beq
M_{h} (r) \!= \!M_{\text{tot}, h} \left(1 + \frac{r_{sh}^2}{r^2}\right)^{-\frac{3}{2}} \!\!, M_{l} (r) \!= \! M_{\text{tot}, l} \left(1 + \frac{r_{sl}^2}{r^2}\right)^{-\frac{3}{2}} \! \!.
\eeq

\subsection{Numerical Implementation}
To implement the two-fluid equations outlined earlier, we first convert the physical quantities $x= (r,\rho,M,\sigma,u\equiv 3 \sigma^2/2,t,L)$ into dimensionless quantities $\hat x \equiv x/x_0$ by introducing a set of fiducial quantities $x_0$. The fiducial quantities are  constructed based on the scale density and radius of the initial profile of  one fluid; here, we choose those of the heavier fluid:
\begin{align}
r_0 ={}& r_{sh}, \quad \rho_0 = \rho_{sh}, \label{eq:r0}\\
M_0 ={}&4\pi r_{0}^3 \rho_{0}, \quad \sigma_0 = r_{0} (4\pi G \rho_{0})^{1/2}, \quad u_0 = \sigma_0^2,  
\label{eq:sigma0}\\
t_{0} ={} & \frac{\sigma_0^3}{12 \pi G^2 m_h \rho_0 \ln \Lambda_{hl} } =   \frac{(4\pi)^{1/2} \rho_{0}^{1/2} r_{0}^3 }{3 G^{1/2}m_h \ln \Lambda_{hl}},  \label{eq:t0} \\
L_0 = {}& \f{G M_0^2}{r_0 t_0} =  3 (4\pi G \rho_0 )^{3/2}  r_{0}^2  m_h \ln \Lambda_{hl}. \label{eq:L0}
\end{align}
Note that the definition of $t_0$ mimics that of the central relaxation time of a one-component stellar cluster~\cite{1987degc.book.....S}, which is given by 
\beq
t_{rc\star} = \frac{0.338 \sigma_{c\star}^3}{G^2 m_\star \rho_{c\star} \ln \Lambda_\star}
\eeq
where $\sigma_{c\star}$, $\rho_{c\star}$, $m_\star$, and $\ln \Lambda_\star$ are the central 1D velocity dispersion, central density, stellar mass, and Coulomb logarithms of the stellar cluster, respectively.

The resulting dimensionless equations are
\begin{align}
 \f{\partial \t M_h}{\partial \t r} ={}& \t \rho_h \t r^2, \quad \f{\partial \t M_l}{\partial \t r} = \t \rho_l \t r^2,\\
\f{\partial}{\partial \t r}\left( \t \rho_h \t \sigma_h^2 \right) ={}& -\frac{ (\t M_l + \t M_h) \t \rho_h }{\t r^2}, \nonumber\\
\f{\partial}{\partial \t r}\left( \t \rho_l \t \sigma_l^2 \right) ={}& -\frac{ (\t M_l + \t M_h) \t \rho_l }{\t r^2},
\label{eq:starfid5}\\
 \frac{D \t u_h}{D \t t} ={}& - \frac{\partial \t L_h}{\partial \t M_h} - c_1  \frac{\t \rho_l  \left(\t u_h- \frac{m_l}{m_h} \t u_l\right)}{(\t u_h +\t u_l)^{3/2}}, \nonumber\\
\f{D \t u_l}{D \t t} ={}&  -\frac{\partial \t L_l}{\partial \t M_l}  + c_1  \frac{\t \rho_h  \left(\t u_h- \frac{m_l}{m_h} \t u_l\right)}{(\t u_h +\t u_l)^{3/2}},
\label{eq:starfid6}\\
\t L_h=   {}& - c_2 \frac{\ln \Lambda_h}{\ln \Lambda_{hl}} \t \rho_h \t r^2 \t u_h^{-1/2} \f{\partial \t u_h}{\partial \t r}, \nonumber\\
 \t L_l  =  {}& -c_2 \frac{m_l}{m_h} \frac{\ln \Lambda_l}{\ln \Lambda_{hl}}  \t \rho_l \t r^2 \t u_l^{-1/2} \f{\partial \t u_l}{\partial \t r},
 \label{eq:starfid7}
\end{align}
where $c_1 = (3\pi)^{-1/2} = 0.326$, and $c_2 = 2^{1/2} \alpha b/9 =0.086$. Note that in deriving~\eqref{starfid6} from~\eqsref{4_1}{4_2}, we have assumed that the densities, $\rho_l$ and $\rho_h$, remain unchanged during the heat conduction and dynamical heating/cooling step. They will be changed and the velocity updated during the relaxation step, thereby implementing an ``operator splitting" technique (see more details below). Given the mass hierarchy~\eqref{hierarchy},  \eqsref{starfid6}{starfid7} can be simplified into
\begin{align}
 \frac{D \t u_h}{D \t t} ={}& - \frac{\partial \t L_h}{\partial \t M_h} - c_1  \frac{\t \rho_l  \t u_h}{(\t u_h +\t u_l)^{3/2}}, \label{eq:limit_eq1}\\
 \f{D\t u_l}{D\t t} ={}&   c_1  \frac{\t \rho_h  \t u_h}{(\t u_h +\t u_l)^{3/2}},\\
\t L_h =  {}&- c_2 \t \rho_h \t r^2 \t u_h^{-1/2} \f{\partial \t u_h}{\partial \t r}, \label{eq:limit_eq3}\\
\hat L_l ={}& 0.\\
(m_h \gg{}&  m_l) \nonumber
\end{align}

We solve the set of dimensionless equations, together with the boundary conditions (see~\secref{bc}) and initial condition (see~\secref{ic}), by extending the ``conduction/interaction--relaxation" iterative methodology~\cite{Pollack:2014rja, Essig:2018pzq} with one more step, ``realignment''. 

We first discretize the two-component spherical systems into two sets of $N$ spherical zones, one for the heavier fluid and one for the lighter fluid. Initially, the spherical zones have logarithmically evenly spaced boundaries, $\{\hat r_1,\cdots, \hat r_N\}$.  During the simulation, we attribute the extensive quantities,  $\hat M_{h(l)}$ and $\hat L_{h(l)}$, to the boundaries of the Lagrangian zones and the intensive quantities, $\hat \rho_{h(l)}$, 
$\hat \sigma_{h(l)}$ and $\hat{u}_{h(l)}$ to the mean radii between two adjacent boundaries. 

The simulation starts with a ``conduction/interaction'' step. In this step, the change in the specific kinetic energy, $\Delta \hat u_{h(l),j}$\footnote{Note that the comma in the subscript is not a derivative symbol.},  for the $j$-th zone of heavier (lighter) fluid is derived from two sources: (1) contributions from adjacent zones, $(j-1)$-th and $(j+1)$-th zones, due to conduction within its own fluid, and (2) contributions from the corresponding $j$-th zone of the other fluid due to dynamical heating/cooling. We set the time interval, $\Delta \hat t$,  small enough to ensure that the maximum specific kinetic energy change rate 
\beq
\epsilon = \max\left\{\frac{|\Delta \hat u_{h,j}|}{\hat u_{h,j}}, \frac{|\Delta \hat u_{l,j}|}{\hat u_{l,j}}\right\},\quad  j=1,\ldots,N
\eeq does not exceed a threshold $\epsilon_\text{thr} = 10^{-3}$. During this step, the densities of all the zones of the two fluids, $\hat \rho_{h(l),i}$, remain the same.

One needs to pay attention to a technical requirement for numerically solving the equation for the conduction/interaction step. Consider the limit $m_l\ll m_h \ll M_\text{tot}$. Within this limit,~\eqsref{limit_eq1}{limit_eq3} can be combined into
\begin{align}
 \frac{D\t u_h}{D\t t} ={}&  \f{2 c_2}{\t \rho_h \t r^2}\frac{\partial}{ \partial \t r} \left( \t \rho_h \t r^2 \frac{\partial \sqrt{\t u_h}}{\partial \t r}\right) -\f{c_1 \t \rho_l \t u_h }{(\t u_h+ \t u_l)^{3/2}} \nonumber\\
={}& \f{2 c_2}{\t r^2}\left[\left(1+\f{\partial \ln \t \rho_h}{\partial \ln \hat r}\right)\f{\partial \sqrt{\t u_h}}{\partial \ln \t r}+\f{\partial^2 \sqrt{\t u_h}}{(\partial \ln \t r)^2}\right] \nonumber\\
&-\f{c_1 \t \rho_l \t u_h  }{(\t u_h+ \t u_l)^{3/2}},
\label{eq:limit_eq1_eq3}
\end{align}
The discretized  version of~\eqref{limit_eq1_eq3} is a function of the spatial step $\Delta \ln \hat r$ and the time step $\Delta \hat t$. Solving the discretized equation with an explicit algorithm follows the general formula
\beq
\frac{\hat u^{n+1}_{h} - \hat u^n_{h}}{\Delta t} = F (\hat u^n_h, \Delta \ln \hat r),
\label{eq:form_1}
\eeq
where $F$ represents the discretized expression of the right-hand side of~\eqref{limit_eq1_eq3} and the superscript on the variable represents the number of time steps. 

In order for~\eqref{form_1} to  converge, the CFL condition must be satisfied, i.e., the discrete time step ($\Delta \hat t$) must be smaller than the time for a wave ($\hat u$) to travel between the discrete spatial step ($\Delta \ln \hat r$). It is straightforward to check that~\eqref{form_1} does not always satisfy the CFL condition for reasonable choices of  $\Delta \hat t$, given 
sufficiently small $\Delta \ln \hat r$. Special attention is required in the discretization, which becomes complicated since  $\Delta \hat t$ and $\Delta \ln \hat r$ are dynamical.  To better handle the CFL condition, we replace the explicit algorithm with a semi-implicit algorithm to solve the discretized~\eqref{limit_eq1_eq3}
\beq
\frac{\hat u^{n+1}_{h} - \hat u^n_{h}}{\Delta t} = F (\hat u^{n+1}_h, \hat u^n_h, \Delta \ln \hat r).
\label{eq:form_2}
\eeq
Inside $F$, some of the original explicit variables $\hat u^n_h$ are replaced with implicit variables $\hat u^{n+1}_h$. The replacement is chosen such that the dependence of $\hat u^{n+1}_h$ stays at the linear order. We prove 
in~\appref{CFL} that under the semi-implicit algorithm, for arbitrary choice of $\Delta t$ and $\Delta \ln \hat r$, solving the discretized~\eqref{limit_eq1_eq3}  always satisfies the CFL condition and is thus unconditionally stable.

Following the conduction/interaction step, the system typically deviates from the hydrostatic equilibrium. To return to the hydrostatic equilibrium, we implement the ``relaxation'' step  for the Lagrangian zones of the heavier and the lighter fluid separately. The relaxation is executed to the linear order in radii changes based on the hydrostatic equations~(\ref{eq:21}--\ref{eq:22}) while maintaining constant (to the linear order) enclosed mass, $\hat M_{h(l), j} - \hat M_{h(l), j-1}$, and specific entropy, $\hat p_{h(l),j} \hat \rho_{h(l),j}^{-5/3}$, for each fluid zone, where $\hat p_{h(l),j} \equiv \hat \rho_{h(l),j} \hat \sigma_{h(l),j}^2$. After relaxation, the radii, the densities, and the specific kinetic energies of the zones are updated.

Given the radii changes, the Lagrangian zones with the same subscript $j$ for the heavier and lighter fluid no longer align with each other. The misalignment prevents us from performing a subsequent step of ``conduction/interaction'' for the zones with the same $j$. We introduce a \emph{new} ``realignment" step to overcome the issue. We interpolate the relaxed density profiles of each fluid using a common set of logarithmically evenly spaced radii, ranging from  $\hat r_1 = \min\{\hat r_{h,1}, \hat r_{l,1}\}$ to $\hat r_N = \max\{\hat r_{h,N}, \hat r_{l,N}\}$. At the inner boundary  $\hat r_1 = \min\{\hat r_{h,1}, \hat r_{l,1}\}$, extrapolation is required for the density  profile of the outward-moving fluid. In such cases, we impose the regularity conditions given by Eq.~(\ref{eq:reg1}-\ref{eq:reg2}), i.e., the extrapolated values are set equal to the innermost interpolated values of the density profile. The realigned enclosed mass profiles and 1D velocity dispersion profiles are consequently derived.

Once realigned, we go back to the conduction/interaction step and repeat the entire ``conduction/interaction--relaxation--realignment'' sequence to  track the stellar heating process continuously. The simulation concludes when the innermost density of one fluid attains a predetermined large value, specifically $\max\{\hat \rho_{h,1}, \hat \rho_{l,1}\} \gtrsim 10^8 $. The innermost density of the collapsed fluid grows super-exponentially fast near the end of the gravothermal evolution. We call such an end phase the ``gravothermal catastrophe". In practice, we use the conclusion time of the simulation to approximate the time that a gravothermal catastrophe occurs ($\hat t_c$). 

Note that N-body simulations have demonstrated that the gravothermally collapsed GCs could form tightly bounded binaries. Those binaries will act as heat source of the stellar cluster through three-body interactions, reversing the gravothermal collapse and driving gravothermal expansion, which could subsequently lead to gravothermal oscillations~\cite{1984ApJ...277L..45C, 1985IAUS..113..139H}.
The presence of a massive, central black hole can
also reverse core collapse~\cite{1977ApJ...217..281S, Shapiro:2018vju}. We will re-explore such possibilities in a future study.

\begin{figure*}[t]
   \centering
   \includegraphics[width=0.96\textwidth]{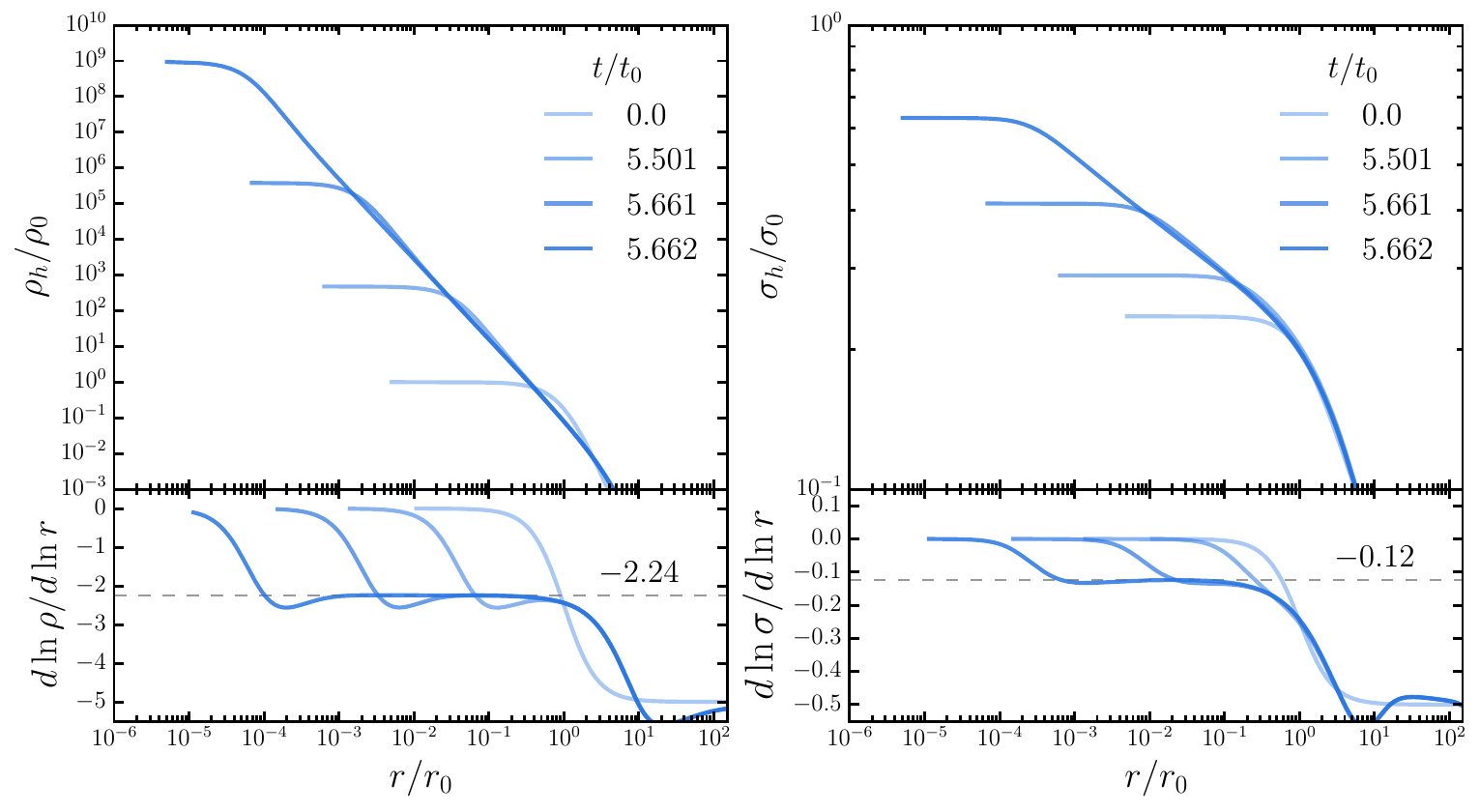} 
   \caption{Snapshots of density (upper left) and 1D velocity dispersion (upper right) profiles for the two-fluid simulation in the one-fluid limit. The lower left and right panels show the log-slopes of the density and 1D velocity dispersion profiles, respectively.}
   \label{fig:one-fluid}
\end{figure*}

\section{Validation}
\label{sec:validation}

Before applying our semi-implicit, fluid conduction code to evolve astrophysically interesting, two-fluid systems, we want to test its reliability. We do this in two stages. We first compare our simulation results for a one-fluid case with results from previous one-fluid simulations in~\secref{check1}. Then, in~\secref{check2}, we perform a self-consistency check by splitting the one-fluid into two pieces, running our two-fluid code on the
combined system and checking that the results are identical to the one-fluid implementation of the same system.

\subsection{Simulations in the one-fluid limit}
\label{sec:check1}

By ``one-fluid", we mean that one of the two fluids, for example, the heavier fluid, makes up 100\% of the mass fraction of the self-gravitating system. This limit can be achieved by taking $\xi  \to 0$ in our two-fluid setup. In such a scenario, the set of eight fluid equations is simplified to 
\begin{align}
&\f{\partial}{{\partial r}}{M_h}= 4\pi r^2 \rho_h, \quad \f{\partial}{\partial r}\left( \rho_h \sigma_h^2 \right) = -\frac{G M_h \rho_h }{r^2} \nonumber \\
&\rho_h \sigma_h^2 \left(\f{D}{D t}\right) \ln\f{\sigma_h^3}{\rho_h} = -\frac{1}{4\pi r^2}\frac{\partial L_h}{\partial r}, \nonumber \\
& \frac{L_h}{4\pi r^2} = - \frac{\alpha b G  \rho_h m_h \ln \Lambda_h}{\sqrt{3} \sigma_h}   \frac{\partial \sigma^2_h}{\partial r} .
\label{eq:fluid_single}
\end{align}
In our actual simulation, we take $\xi = 10^{-6}$, $\zeta = 1$, and $N=150$ Lagrangian zones for both fluids with $\hat r_{1} = 10^{-2}$ and $\hat r_\text{out} = 10^3$ at $\hat t = 0$.

The resulting density and the 1D velocity dispersion profiles are shown in~\figref{one-fluid}. Table~\ref{tab:tab1} summarizes the key quantities of the simulation: the collapse time, $\hat t_c$, and the log-slopes for the inner part of the collapsed density and 1D velocity dispersion profiles. The results of the two-fluid simulation in the one-fluid limit are in reasonable agreement with those of the previous one-fluid simulation reported in~\cite{Shapiro:2018vju}. (Details for the collapse time estimation and notation translation can be found in~\appref{trans}.) Note that although the resulting density profile of the collapsed fluid is slightly steeper than that in~\cite{Shapiro:2018vju}, the log-slopes of the inner part of the profiles still satisfy the relation~\cite{Shapiro:2018vju}
\beq
\left.\f{d \ln \sigma_h}{d \ln r} \right|_c =  \left.\frac{1}{2} \f{d \ln \rho_h}{d \ln r}\right |_c + 1.
\label{eq:twoslope}
\eeq
The relation~\eqref{twoslope} can be derived by assuming the density and 1D dispersion profiles follow the single-power laws, $\rho_h \propto r^x$ and $\sigma_h \propto r^y$. Substituting this ansatz into the hydrostatic equations (first 
line of~\eqref{fluid_single}), one finds that the LHS and RHS scale as $r^{x+2y-1}$ and $r^{2x+1}$, respectively. Requiring the two sides to follow the same scaling yields $y=x/2+1$.

\begin{table}[htbp]
   \centering
 \topcaption{Comparing the collapse time, the log-slope of the inner part of the collapsed density profile, and the log-slope of the inner part of the collapsed 1D velocity dispersion profile of this work and~\cite{Shapiro:2018vju}.}
   \label{tab:tab1}
   \begin{tabular}{ cc@{\hskip 15pt}c} 
      \hline
          & This Work & Ref. \cite{Shapiro:2018vju}\\
      \hline
     $\hat t_c \equiv t_c/t_0$      & 5.662 & $\gtrsim 5.542$\\
      $\left.\frac{d \ln \rho_h}{d \ln r}\right|_c$       &  $-2.24$  & $-2.21$ \\
      $\left.\frac{d \ln \sigma_h}{d \ln r}\right|_c$       &  $-0.12$  & $-0.11$ \\
\hline
   \end{tabular}
  \end{table}

\begin{figure*}[t]
   \centering
   \includegraphics[width=0.96\textwidth]{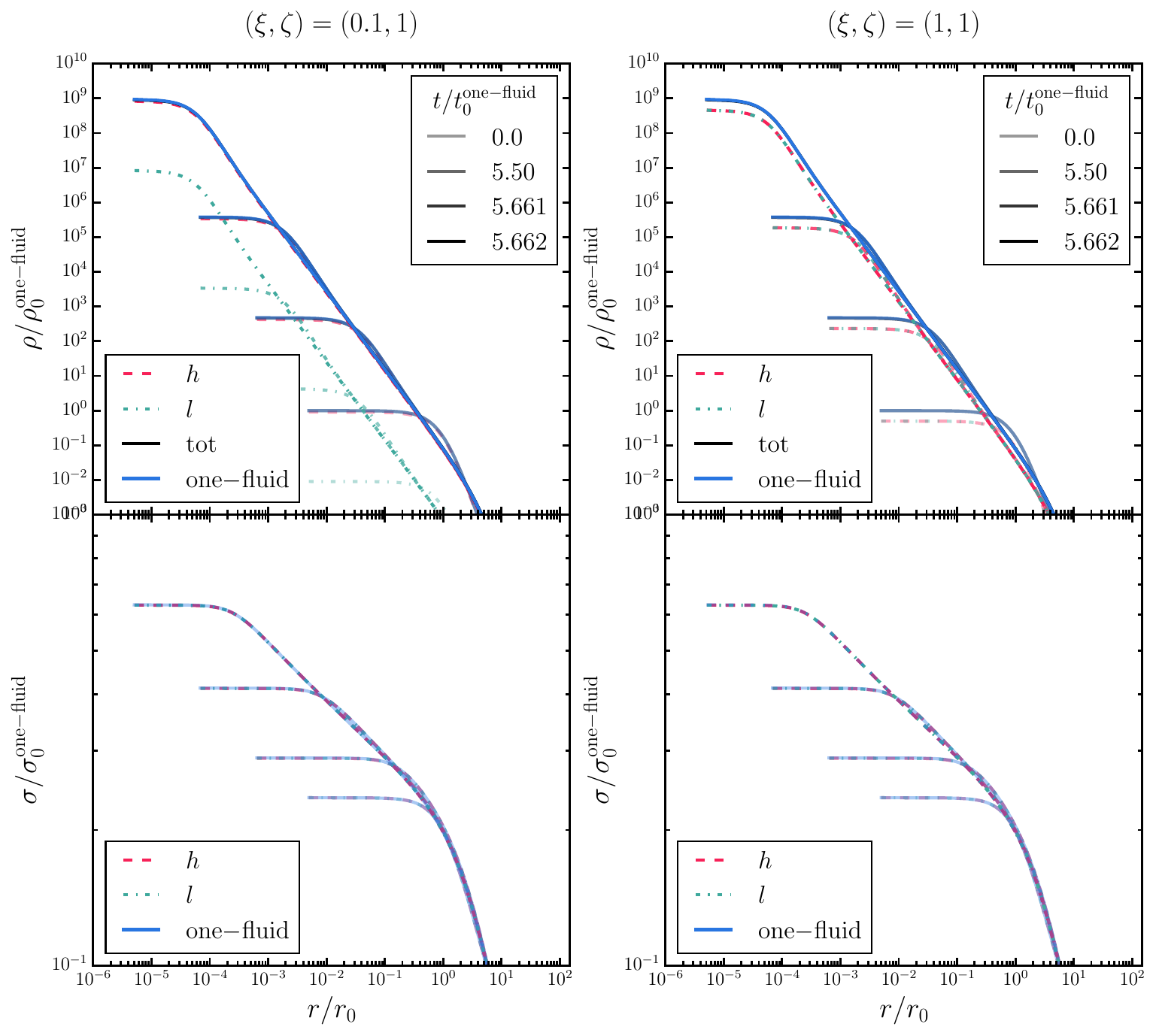}   
    \caption{Snapshots of 
 the density profiles (upper row) and the 1D velocity dispersion profiles (lower row) for the one-fluid-splitting-in-two simulations at 
 four selected times during the evolution. These snapshots are obtained from simulations conducted for the scenarios $(\xi, \zeta) = (0.1,1)$ (left column) and $(\xi, \zeta) = (1,1)$ (right column), respectively. For each simulation, the densities, 1D velocity dispersions, and evolution time are normalized with respect to the fiducial density,  fiducial 1D velocity dispersion, and  fiducial time of the one-fluid limit, respectively, shown in blue lines. The red-dashed (green-dash-dotted) lines denote the profiles corresponding to the heavier (lighter) fluid, while the black lines in the upper row represent the total density of two fluids and are essentially indistinguishable from the blue lines. In all the panels, the one-fluid limit (blue line) profiles are identical to those presented in~\figref{one-fluid}.
}
   \label{fig:twofluid_equal}   
\end{figure*}

\subsection{Simulation of one-fluid splitting in two}
\label{sec:check2}

Next, we perform a self-consistency test of the two-fluid code by splitting one fluid into two components and checking that the two-fluid evolution is unchanged from the one-fluid evolution. The designated two fluids are artificially labeled `$h$' and `$l$', which no longer stand for heavier and lighter fluids. We set $\zeta=1$, $m_l/m_h = 1$, and $\ln \Lambda_h = \ln \Lambda_l$. In such a scenario, the velocity dispersions are identical
and should remain so. Hence, the dynamical heating/cooling term $R$, ~\eqref{dyn}, should vanish, and the fluid equations for the $`h'$ and $`l'$ components are identical and should reduce to~\eqref{fluid_single} for each species.
 
We check two scenarios of artificial splitting: $(\xi, \zeta)  = (0.1, 1)$ and $(\xi,\zeta )=(1, 1)$. The setup of the Lagrangian zones for the two-fluid cases is identical to that adopted in the one-fluid limit scenario described in~\secref{check1}.

The upper (lower) row of~\figref{twofluid_equal} shows snapshots of the density (1D velocity dispersion) profiles for the two scenarios. All densities, 1D velocity dispersions, and evolution times are normalized by  the corresponding fiducial values of the one-fluid limit (indicated by subscript ``one-fluid"). 

For the scenario $(\xi, \zeta)  = (0.1, 1)$, the heavier (red-dotted lines) and lighter fluid (green dash-dotted lines) maintain a fixed density ratio of 10 as they collapse, whereas their 1D velocity dispersion profiles evolve identically. This behavior follows from~\eqref{starfid5} for the initial condition and the absence of heat exchange discussed earlier. For the scenario $(\xi, \zeta) = (1,1)$ , both fluids exhibit identical evolution in their density profiles and 1D velocity dispersion profiles.

In addition, the total density from the heavier and lighter fluids (black solid lines, underneath the blue solid lines) is shown in the upper row, alongside the density and 1D velocity dispersion profiles (blue solid lines) from the one-fluid limit at approximately the same evolution time. The sum of the two-fluid densities closely matches the one-fluid limit, and the heavier (or lighter) fluid’s 1D velocity dispersion also evolves in the same way as that of the one-fluid limit.

In summary, the above agreement, both in the density and the velocity dispersion profiles, between the one-fluid-splitting-in-two scenario and the one-fluid scenario demonstrates that our code is self-consistent.

\section{Results}
\label{sec:results}

\subsection{Comparing the two-fluid simulation and the N-body simulation}
\label{sec:check3}

\begin{figure*}[t!]
	\includegraphics[width=0.96\textwidth]{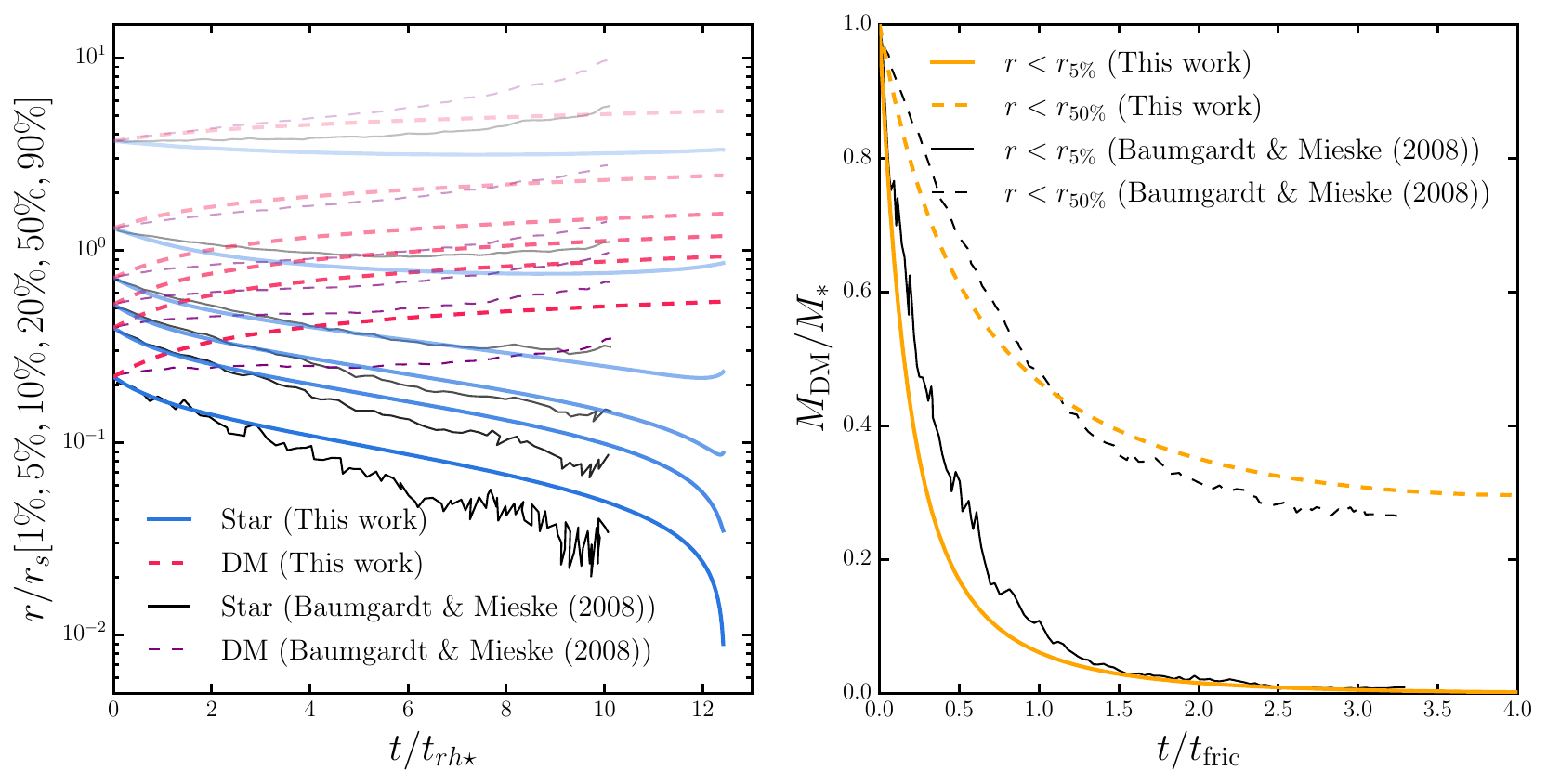}
    \caption{Comparison of the evolution curves of benchmark I of~\tabref{baumgardt_bm} for fluid vs. N-body simulations. ({Left}) The blue (black) solid and red (purple) dashed lines represent the evolution of the radii of Lagrangian zones of the stars and DM from the fluid (N-body) simulations, respectively.  From bottom to top, the gradually lighter lines show the evolution for zones enclosing 1\%, 5\%, 20\%, 50\%, and 90\% of the total mass of stars or DM. The radii are normalized to the scale radius of the initial Plummer profile $r_s$, and the time is normalized to the relaxation time of the stars $t_{rh\star}$. Densities from the N-body simulations~\cite{Baumgardt:2008abc} are rescaled by a factor of 1.7 to account for normalization differences. ({Right}) The evolution of the DM-to-star mass ratio. Solid (dashed) lines show the ratio evolution within the 5\% (50\%) Lagrangian radii of the total mass. Orange (black) lines represent fluid (N-body) simulation results. Time is normalized to the dynamical friction time of the stars $t_\text{fric}$.
}
    \label{fig:baumgardt_bm}
\end{figure*}

Ref.~\cite{Baumgardt:2008abc} utilizes the N-body simulation to study dynamical heating in a system composed of stars (the heavier component) and ultraheavy DM (the lighter component) with the same initial Plummer profiles.
However, due to the resolution limitations of the N-body method, the N-body evolution does not extend to late times when the stars are  in the deep gravothermal collapse regime. Here, we revisit the evolution of the system using the conduction fluid method and compare our results with those of~\cite{Baumgardt:2008abc}, extending the analysis to later times.

Table~\ref{tab:baumgardt_bm}~presents the parameters for four benchmark systems (as provided in Table~1 of \cite{Baumgardt:2008abc}). These include the total number of the  stars, $N_\star$, and the DM particles, $N_\chi$, the mass of DM N-body particle, $m_\chi$, the total mass ratio between stars and DM, $M_{\text{tot}\star}/M_{\text{tot}\chi}$. Additionally, we include derived parameters in Table~\ref{tab:baumgardt_bm}, such as the average mass of stars, $\vev{m_\star}$, the mass ratio between a DM  particle and the average mass of stars, $m_\chi/\vev{m_\star}$, and the ratios of Coulomb logarithms, ${\ln \Lambda_\star}/{\ln \Lambda_{\star\chi}}$ and ${\ln \Lambda_\chi}/{\ln \Lambda_{\star\chi}}$, with $\gamma=0.11$. It is important to note that Ref.~\cite{Baumgardt:2008abc} models a population of stars sampled from  a continuous mass distribution, which should ideally be treated as a multi-component fluid. In our analysis, we simplify this by treating the stars as a uniform fluid with a single average stellar mass $\vev{m_\star} \equiv M_{\text{tot}\star}/N_\star$. For all benchmark systems, the resulting average stellar mass is $0.335\, \text{M}_\odot$, only slightly smaller than the mean stellar mass of $0.344\, \text{M}_\odot$ mentioned in Sec.~II of~\cite{Baumgardt:2008abc}. For all the simulations, we set $\zeta=1$.

\begin{table}[h]
 \centering
   \topcaption{Parameters adopted for the two-fluid and N-body comparison.}
      \label{tab:baumgardt_bm}
   \begin{tabular}{c|cccc|cccc}
      \hline
    \#& $N_\star$ & $N_\chi$ & $\f{m_\chi}{\text{M}_\odot}$ &$\f{M_{\text{tot}\chi}}{M_{\text{tot}\star}}$ & $\f{\vev{m_\star}}{\text{M}_\odot}$ & $\f{m_\chi}{\vev {m_\star}}$ & $\f{\ln \Lambda_\star}{\ln \Lambda_{\star\chi}}$ & $\f{\ln \Lambda_\chi}{\ln \Lambda_{\star\chi}}$ \\
         \hline
        I& 8224 & 91776 & 0.03 & 1 & 0.335 & 0.09 & 0.95 &  1.10 \\
       II&   4285 & 95715 & 0.03 & 2 & 0.335 & 0.09 & 0.95  & 1.10 \\
       III&  8223 & 27528 & 0.1 & 1 & 0.335 & 0.3 & 0.99  & 1.00 \\
       IV&  8224 & 183576 & 0.015 & 1 & 0.335 & 0.045 & 0.94  &  1.16\\
         \hline
   \end{tabular}
\end{table}

The left panel of~\figref{baumgardt_bm} presents the temporal evolution of the Lagrangian radii for benchmark I from~\tabref{baumgardt_bm}, based on the fluid simulations. The blue solid (red dashed) lines, with progressively lighter shades, represent the radii of stellar (or DM) fluids that enclose 1\%, 5\%, 20\%, 50\%, and 90\% of the total stellar mass, $M_{\text{tot}\star}$, (or the total DM mass, $M_{\text{tot}\chi}$). These radii are normalized to the scale radius of the initial Plummer profile for stars (or DM), $r_s=r_{s\star} = r_{s\chi}$. The evolution time is normalized to the stellar half-mass relaxation time, $t_{rh\star}$,  used in~\cite{1987degc.book.....S, Baumgardt:2008abc}:
\beq
t_{rh\star} = 0.138 \f{M_{\text{tot}\star}^{1/2}r_{50\%\star}^{3/2}}{G^{1/2} m_\star \ln \Lambda_\star} = 0.356 \frac{\ln \Lambda_{\star\chi}}{\ln \Lambda_\star} t_0.
\label{eq:time_halfmass}
\eeq
The second equality uses~\eqref{t0}, and the stellar half-mass radius, $r_{50\%\star}$, is set to the value of the initial Plummer profile. For comparison, the N-body simulation results from Fig.~2 of~\cite{Baumgardt:2008abc} are shown as black solid (purple dashed) lines for the stellar (DM) Lagrangian radii. Since the two sets of simulations use different normalization schemes for the radii, the radii from~\cite{Baumgardt:2008abc} have been multiplied by a factor of 1.7 to align with the fluid simulations at $t=0$.

\begin{table*}[t]
 \centering
   \topcaption{The mass ratios of DM to stars inside the 5\% Lagrangian radius, $r_{5\%}$, and the half-mass radius, $r_{50\%}$, of the whole system at $t= t_{rh\star}, 2t_{rh\star}, 10t_{rh\star}$. Values in parentheses are results from~\cite{Baumgardt:2008abc}.}
      \label{tab:baumgardt_fdm}
     \begin{tabular}[t]{c|cc@{\hskip 15pt}cc@{\hskip 15pt}cc}
      \hline
      &    \multicolumn{2}{c}{$M_\chi/M_\star|_{t = t_{rh\star}}$} &   \multicolumn{2}{c}{$M_\chi/M_\star|_{t = 2 t_{rh\star}}$}  &   \multicolumn{2}{c}{$M_\chi/M_\star|_{t =10  t_{rh\star}}$} \\
      \#& $r<r_{5\%}$ & $r<r_{50\%}$ & $r<r_{5\%}$ & $r<r_{50\%}$ & $r<r_{5\%}$ & $r<r_{50\%}$ \\ 
      \hline
      I& 0.24 (0.39) & 0.68 (0.78) &  0.10 (0.22) & 0.53 (0.63) & 0.002 (0.01) & 0.29 (0.26)\\
      II&   0.56 (0.57) & 1.44 (1.51) & 0.23 (0.29) & 1.16 (1.27) & 0.003 (0.03) &  0.73 (0.55)\\
      III&   0.35 (0.38) & 0.74 (0.81) & 0.19 (0.20) & 0.61 (0.70) & 0.02 (0.01) & 0.37 (0.35) \\
      IV&  0.22 (0.40) & 0.67 (0.78) & 0.09 (0.18) & 0.51 (0.61) & 0.001 (0.01) & 0.29 (0.27) \\
      \hline
   \end{tabular}
\end{table*}

Overall, the conduction fluid and N-body simulations show reasonable agreement, with the former method shrinking the radii over a longer timescale. Both simulations exhibit segregation between the two fluids --  the stars contract inward while DM expands outward.  The left panel of~\figref{baumgardt_bm} shows that the stellar radii from the fluid simulations (blue solid lines) are in good agreement with the N-body results (black solid lines), except for the 50\% and 90\% Lagrangian radii, where the fluid simulations show a mild collapse, while the N-body simulations display slower collapse or slight expansion.
The opposite trend is seen for the DM radii (dashed lines). For smaller Lagrangian radii (1\%, 5\%, 10\%, and 20\%), DM expands more slowly in the N-body simulations compared to the fluid simulations. However, for larger radii (50\% and 90\%), both simulations are in reasonable agreement.

The right panel of~\figref{baumgardt_bm} shows the evolution of the DM-to-star mass ratio for benchmark I, comparing conduction fluid (orange) and N-body (black) simulations. Two regions are considered: one  within the 5\% Lagrangian radius (solid lines) and the other within the 50\% Lagrangian radius (dashed lines) encompassing the total mass. Time is normalized to the ``dynamical friction time" of the stellar component (with a Plummer profile) to match the convention used in Fig.~3 of~\cite{Baumgardt:2008abc}. The relationship between $t_\text{fric}$ and the fiducial time $t_0$~\eqref{t0} is expressed as:
\beq
t_\text{fric} = 3.01 \sqrt{  \frac{M_\text{tot}}{M_{\text{tot}\star}} } t_{rh\star} = 1.08\sqrt{  \frac{M_\text{tot}}{M_{\text{tot}\star}} }\frac{\ln \Lambda_{\star\chi}}{\ln \Lambda_\star} t_0.
\eeq
The results again show a reasonable agreement between the conduction fluid and N-body simulations.

~\tabref{baumgardt_fdm} summarizes the DM-to-star mass ratio from the fluid simulations within two Lagrangian radii ($r_{5\%}$ and $r_{50\%}$) of the total mass at three different time points ($t = t_{rh\star}, 2t_{rh\star}, 10t_{rh\star}$) across all benchmarks. Values in parentheses represent N-body simulation results from\cite{Baumgardt:2008abc}. Similar to the evolution curves in~\figref{baumgardt_bm}, the mass ratios from both simulation methods follow comparable trends as the DM and star segregate over time.

\begin{figure*}[t]
	\includegraphics[width=0.96\textwidth]{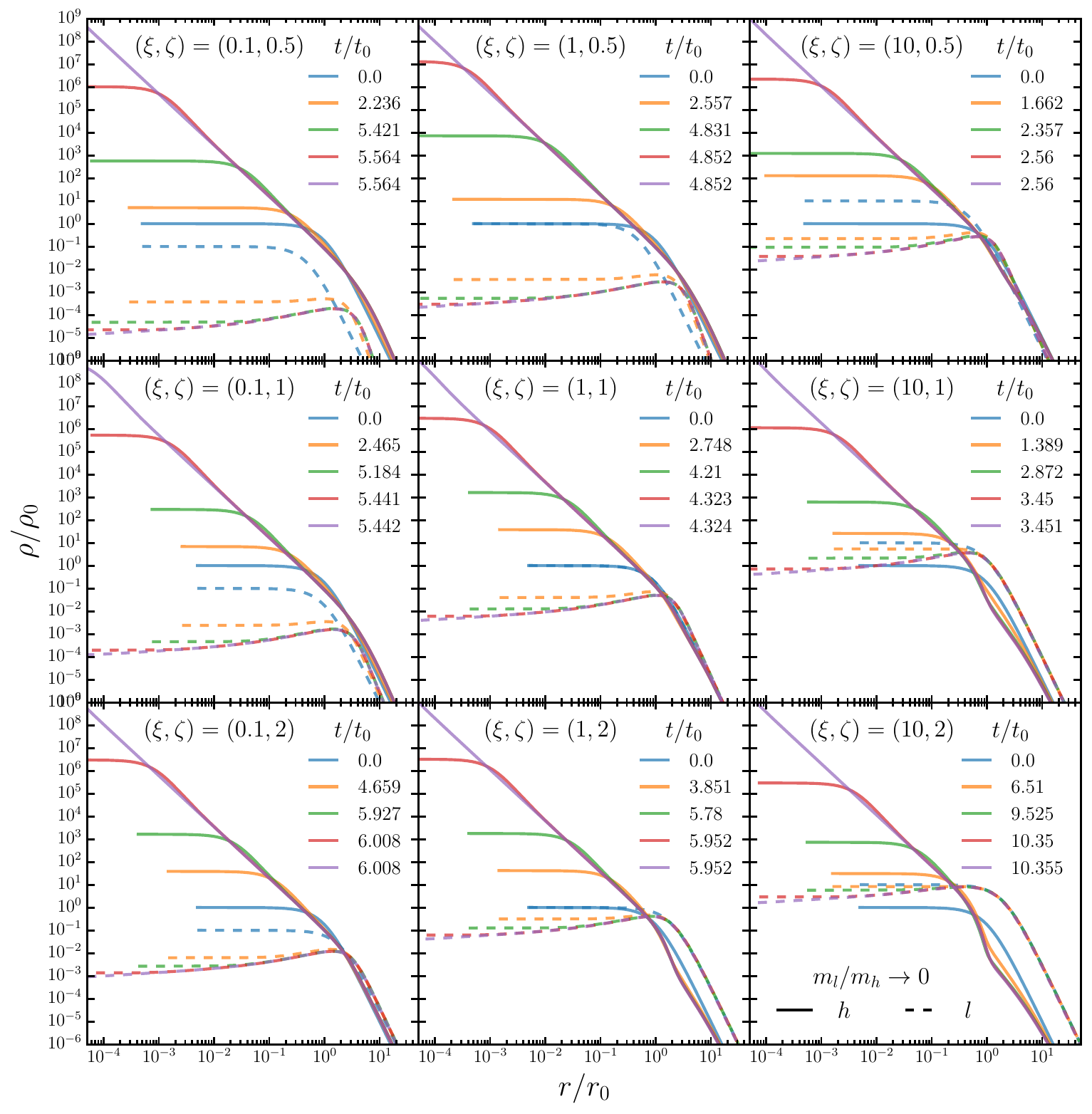}
    \caption{Snapshots of density profiles for heavier (solid lines) and lighter (dashed lines) fluids in the limit $m_l /m_h \to 0$ for $\xi \equiv \rho_{sl}/\rho_{sh}$ to be 0.1 (left), 1 (central), and 10 (right) and $\zeta \equiv r_{sl}/r_{sh}$ to be 0.5 (upper), 1 (middle), and 2 (lower). The evolution time is normalized to the simulation fiducial time $t_0$, which is constructed from the initial scale radius and density of the heavier fluid ($r_0 = r_{sh}$, $\rho_0 = \rho_{sh}$) by~\eqref{t0}.
    }
    \label{fig:pdm}
\end{figure*}

\begin{figure*}[t]
	\includegraphics[width=0.96\textwidth]{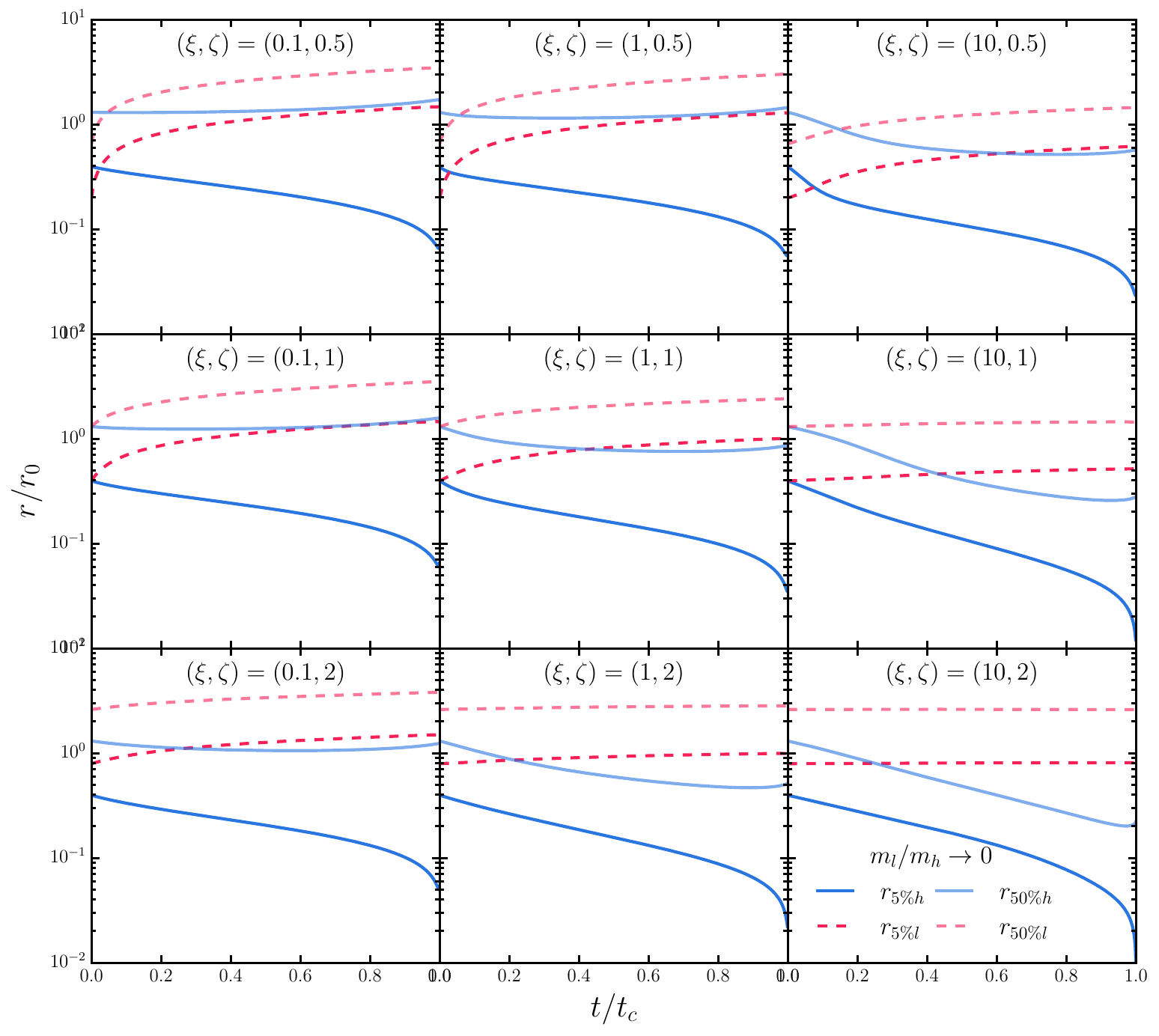}
    \caption{Temporal evolution of the Lagrangian radii  for cases with $\xi = 0.1$ (left), 1 (central), and 10 (right) and $\zeta = 0.5$ (upper), 1 (middle), 2 (lower) in the limit $m_l /m_h \to 0$. For each panel, the evolution time for each scenario is normalized to its collapse time.  Lines with darker and lighter solid blues (dashed reds)  represent the 5\% and 50\% Lagrangian radii of the heavier (lighter) fluid, respectively. 
  }
    \label{fig:pdmB}
\end{figure*}

\begin{figure*}[t]
	\includegraphics[width=0.96\textwidth]{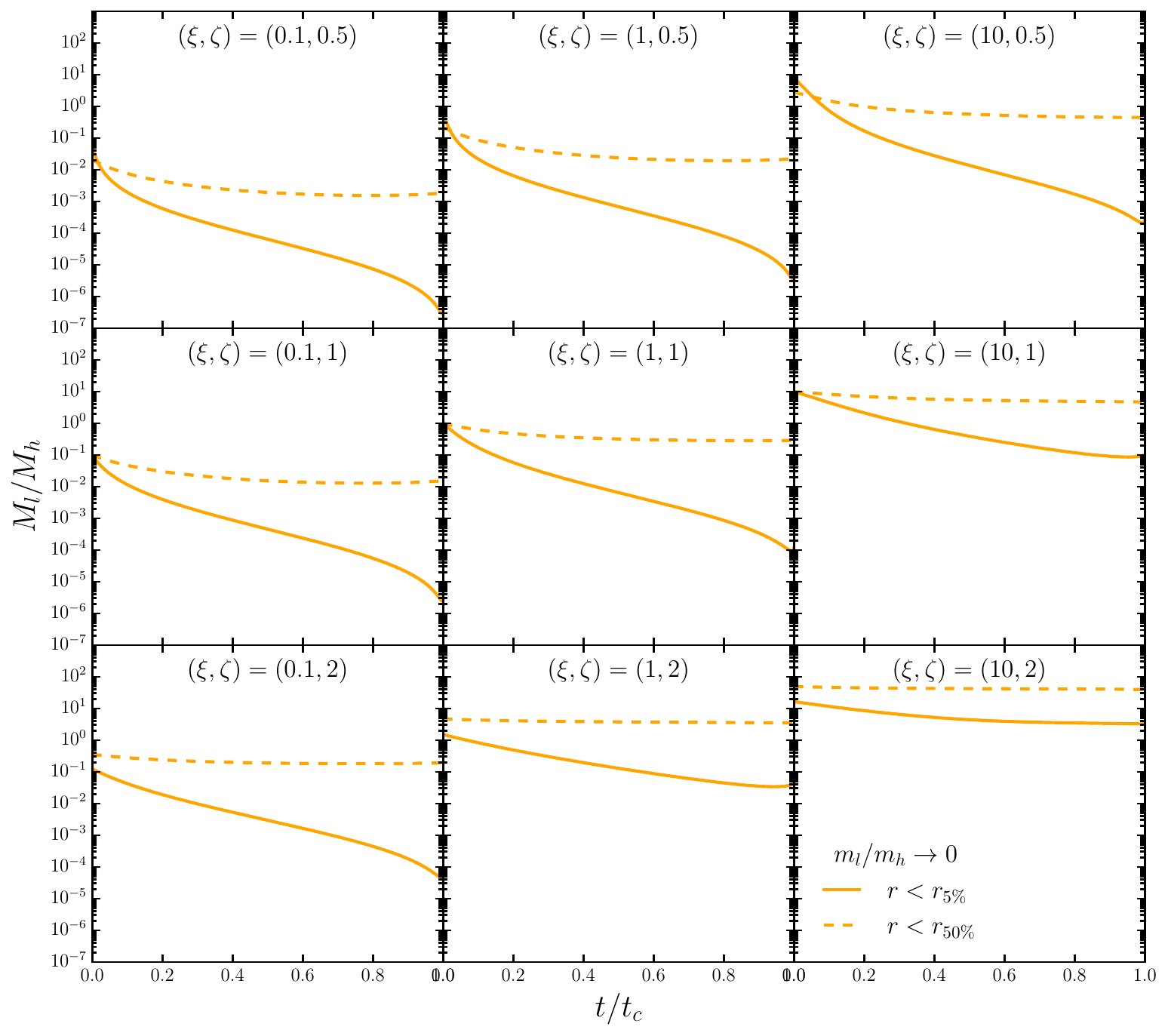}
    \caption{Temporal evolution of lighter-to-heavier fluid mass ratio with $\xi = 0.1$ (left), 1 (central), and 10 (right) and $\zeta = 0.5$ (upper), 1 (middle), 2 (lower) in the limit $m_l /m_h \to 0$. For each panel, the solid and dashed lines show the ratio within 5\% and 50\% Lagrangian radii of the total mass, respectively. The evolution time is again normalized to each collapse time.
  }
    \label{fig:pdmC}
\end{figure*}

\subsection{Two-fluid simulations in the limit $m_h \gg m_l$}
\label{sec:two-fluid-hierarchy}

Ultra-heavy DM candidates, such as PBHs, are well-motivated and can be effectively probed via astrophysical observations. As a specific example, constraints from microlensing, gravitational waves, and galactic dynamics have ruled out PBHs as 100\% of DM for $m \gtrsim 10^{-10}\,\msun$. Nevertheless, there are still mass windows where PBHs can contribute a non-negligible fraction of the DM. For instance, around $m \simeq 10^{-6}\,\msun$, PBHs can make up  $\sim 10\%$ of DM, and around $m \simeq 20\,\msun$, they can similarly account for  $\sim 10\%$. See the review by \cite{Carr:2021bzv}.

Here, we are interested in considering a progenitor galaxy made with stars and PBHs in one of the mass windows mentioned above. In either case, we can treat the mass of the heavier component to be much greater than that of the lighter one, e.g.,  \eqref{hierarchy}.  Given the hierarchy, we consider nine cases with different scale density ratios ($\xi = 0.1, 1, 10$)  and different scale radius ratios ($\zeta = 0.5, 1, 2$) of the two fluids.

~\figref{pdm} shows the resulting snapshots of the density profiles of the heavier and lighter fluids for the nine cases where the radius, density, and time are normalized to the initial scale radius, density, and the fiducial time (\eqsref{r0}{t0}), respectively. In all scenarios, the heavier fluid undergoes collapse while the lighter fluid expands at a certain degree. The core collapses occur at $ t_c = \mathcal{O}(1) t_0 $, where $t_0$ is the fiducial time based on the heavier fluid. The specific values of $t_c/t_0$ are shown in~\tabref{benchmark}.

During the evolution, the most significant density changes  for the two fluids occur in the inner halo.  As $\xi$ or $\zeta$ increases, the relative reduction of the lighter fluid density in the inner region is reduced, indicating that the dynamical heating by the heavier fluid becomes less significant. \figref{pdm} also shows a notable change in the heavier density slope around $r= r_{sh}$ for the case $(\xi,\zeta) =(10,2)$.

~\figref{pdmB} illustrates the resulting temporal evolution of  the 5\% and 50\% Lagrangian radii of the heavier or lighter fluid mass under different initial setups. The evolution time for each scenario is normalized to the corresponding collapse time $t_c$. Examining the rates of change in the curves reveals an interesting dependence of the mass segregation effect on $\xi$ and $\zeta$. When the heavier fluid dominates the progenitor ($\xi = 0.1$), the heating of the lighter fluid is significant, while the contraction of the heavier fluid is relatively mild during the evolution. Conversely, when the lighter fluid dominates the progenitor ($\xi = 10$), the heating of the lighter fluid is mild, but the contraction of the heavier fluid becomes more substantial. For different $\zeta$ with the same $\xi$, we found that the heating of the lighter component becomes less significant when $\zeta$ is larger while the contraction of the heavier fluid remains similar. Defining $t_s$ as when $ r_{5\%} $ of the lighter fluid crosses $ r_{50\%} $ of the heavier fluid, we found that significant mass segregation occurs at $t_s \approx \mathcal{O}(10\%) t_c$ or $t_s \approx t_0$.  Detailed values can be found in~\tabref{benchmark}.

\begin{table}[h]
   \centering
   \topcaption{The collapse and segregation time for various benchmarks. `--' in the $t_s/t_0$ rows indicates  $r_{50\%h} > r_{5\% l}$ through the evolution. The fiducial times $t_0$ for all the benchmarks are computed for the heavier component.}
   \begin{tabular}{@{} c|ccccccccc @{}} 
      \hline
         $(\xi, \zeta)$  & $(0.1, 0.5)$ & $(0.1,1)$ & $(1,0.5)$ & $(0.1,2)$ & $(1,1)$  \\
      \hline
      $t_c/t_0$      & 5.56 & 5.44 & 4.85 &  6.01 & 4.32  \\
      $t_s/t_0$       &   --  & --   & -- & 1.57 & 1.84 \\
      \hline
       \hline
         $(\xi, \zeta)$  &  $(10,0.5)$ & $(1,2)$ & $(10,1)$ & $(10,2)$ \\
      \hline
      $t_c/t_0$       & 2.56 & 5.95 & 3.45 & 10.4 \\
      $t_s/t_0$        & 1.55 & 1.27 & 1.49 & 2.59\\
      \hline
      
   \end{tabular}
   \label{tab:benchmark}
\end{table}

\figref{pdmC} shows the temporal evolution of the mass ratio $ M_l/M_h $ within the 5\% (solid lines) and 50\% (dashed lines) Lagrangian radii of the total mass under the same setup. A depletion by multiple order-of-magnitude of the lighter fluid occurs within $r_{5\%}$ for halos with $M_{\text{tot},l}/M_{\text{tot},h}\lesssim 1$. However, within $r_{50\%}$, the depletion is much milder, with $M_l/M_h$ changing by only a factor of a few throughout the dynamical evolution.

\section{Comparing the simulation results to the observed data for GCs and dwarf galaxies}
\label{sec:observation}

Given the simulation results presented in~\secref{check1} and~\secref{two-fluid-hierarchy}, it is reasonable to adopt the fiducial timescale of the heavier component, $t_0^h$, defined in~\eqref{t0}, as an approximate indicator of both the  mass‑segregation and core‑collapse times for a broad range of $M_l/M_h$ ratios.\footnote{We add a superscript $h$ to emphasize that $t_{0}$ is evaluated for the heavier component.} We therefore compare $t_{0}^h$ with the ages of GCs and dwarf galaxies in the Local Volume, which are usually estimated to be between $10\,\mathrm{Gyr}$ and $13\,\mathrm{Gyr}$. In the following estimate, we set the mass of an individual star to $m_{\star}=1\,\msun$ for all GCs or dwarfs. When the DM particle mass satisfies $m_{\chi}\ll m_{\star}$, the typical dynamical timescales of the stellar–DM systems are  
\beq
t_0^h =  \frac{(4\pi)^{1/2} \rho_{s\star}^{1/2} r_{s\star}^3 }{3 G^{1/2}m_\star \ln (2\gamma M_\text{tot}/m_\star)}.
\label{eq:t0_GC}
\eeq

We  use data compiled by~\cite{1996AJ....112.1487H, Harris:2010ut} for the GCs and LVDB\footnote{\url{https://github.com/apace7/local_volume_database}} for the dwarf galaxies from Milky Way, M31, and Local Field.\footnote{Note that LVDB includes Harris GCs, along with the newly discovered GC in the Milky Way. However, some of the properties of those GCs, such as their dynamical masses, are not included in the current version of LVDB. Therefore we adopt the original Harris catalog instead, which includes key information such as the stellar half-mass relaxation time, $t_{rh\star}$, and whether the GC appears core-collapsed.}

For a given system, we can relate $t_0^h$ to the stellar half-mass relaxation time $t_{rh\star}$ by~\eqref{time_halfmass},
\begin{align}
t_0^h ={}& 2.81 \frac{ \ln (\gamma M_{\text{tot}\star}/m_\star)}{ \ln (2\gamma M_\text{tot}/m_\star)} t_{rh\star} \nonumber\\
={}&\frac{5.8\times 10^6\, \text{yr}}{\ln (2\gamma M_\text{tot}/m_\star)}  \left(\frac{\msun}{m_\star}\right) \left(\frac{M_{\text{tot}\star}}{\msun}\right)^{1/2}\left(\frac{r_{50\%\star}}{\text{pc}}\right)^{3/2},
\label{eq:t0_1}
\end{align}
 assuming the stars follow a Plummer profile at the onset of the evolution. A key challenge in evaluating $t_0^h$ is that we can only observe properties of the evolved system, e.g., $r_{50\%\star}$, and lack direct information about their initial values. To address this, we make an educated guess that the initial $r_{50\%\star} (t=0)$ is roughly the same as the evolved (observed) value $r_{50\%\star} (t=t_\text{obs})$. This assumption is supported by the simulated benchmarks. As shown in the first two columns of \figref{pdmB}, the half-mass radii of the heavier fluid (stars) contract by at most a factor of 2--3 during evolution in systems with a mass ratio $ M_{\text{tot},\chi}/M_{\text{tot}\star} = \xi \zeta^3 \lesssim 8 $. The half-light radius, rather than the stellar half-mass radius, is often measured from observations. Here, we approximate the half-mass radius $ r_{50\%\star} $ to be the same as the 3D half-light radius $ r_{\frac{1}{2}\text{light}} $, and relate the latter to the 2D projected half-light radius, $R_e$, which is more commonly reported, by
\beq
r_{50\%\star} \approx r_{\frac{1}{2}\text{light}} \approx \frac{4}{3} R_e,
\eeq
where the second approximation is valid for a wide range of spherical systems~\cite{Wolf:2009tu}.

\begin{figure*}[htbp]
   \centering
   \includegraphics[width=0.96\textwidth]{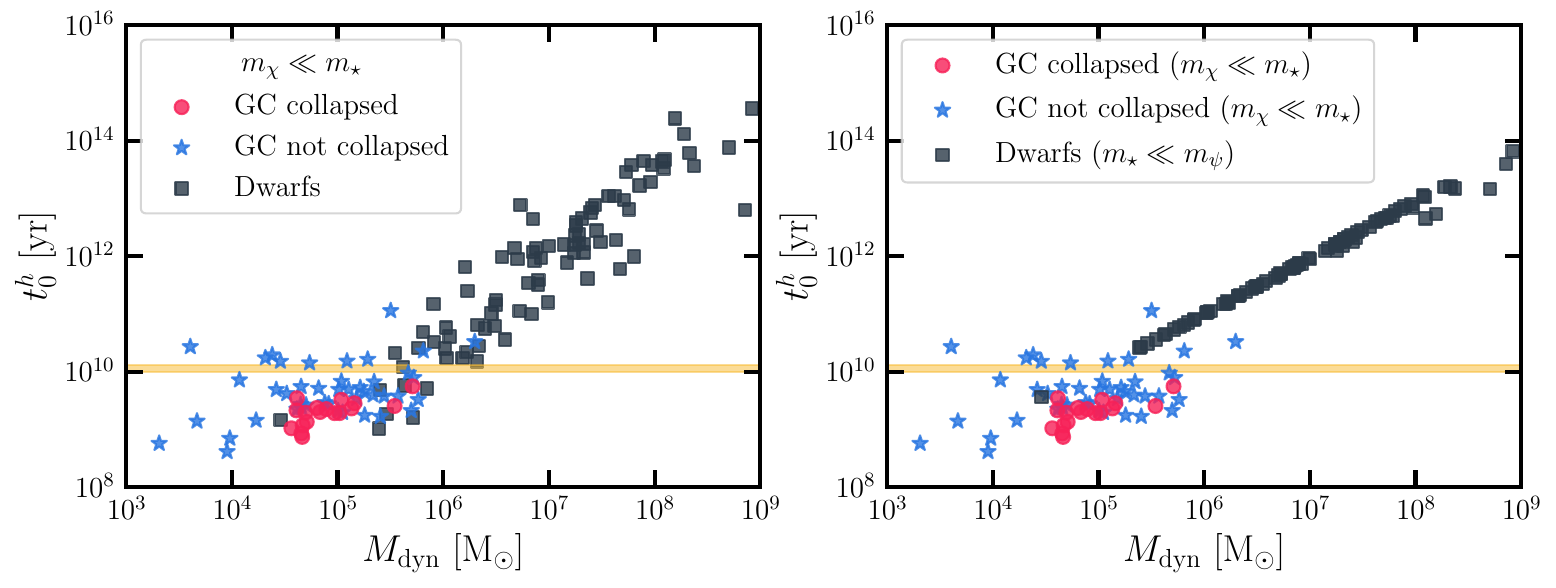} 
   \caption{(Left) Fiducial time of the heavier component $t_0^h$ vs. dynamical mass $M_\text{dyn}$ for Harris  GCs~\cite{1996AJ....112.1487H, Harris:2010ut} and dwarf galaxies compiled by LVDB~\cite{Pace:2024sys} under the assumption $m_\chi \ll m_\star = 1\, \msun$. We set $M_{\text{tot}\chi}/M_\text{tot} \ll 1$ for the GCs and from 0.3 to 1 with a median of 0.98 for the dwarfs. The dots (stars) represent GCs observed with(without) core collapse. The squares represent dwarf galaxies from Milky Way, M31, and Local Field. The yellow band indicates the typical age for those systems (or the time of observation since the onset of the evolution), 10--13 Gyrs.  (Right) A two DM setup ($\psi$ and $\chi$) where we assume $m_\psi = 20\, \msun \gg m_\star \gg m_\chi$. We set
$M_{\text{tot}\psi}/M_\text{tot} \ll M_{\text{tot}\chi}/M_\text{tot} \ll 1$ for GCs (the GC data is the same as the left panel), while for dwarfs we have $M_{\text{tot}\chi}/M_\text{tot} \ll M_{\text{tot}\psi}/M_\text{tot}$ with
$M_{\text{tot}\psi}/M_\text{tot}$ varies from  0.3 to 1
with a median of 0.98. We use the concentration-mass-redshift relation from Ref.~\cite{Ludlow:2016ifl} to estimate the scale radius of the dwarf halos. }
   \label{fig:trh_ML}
\end{figure*}

For the total stellar mass $M_{\text{tot}\star}$, we will assume $M_{\text{tot}\star}/ M_\text{tot} \approx 1$ for the GCs. For the dwarf galaxies, we use  \texttt{mass\_stellar} provided by LVDB, which computes the stellar mass from the V-band absolute magnitude, assuming a mass-to-light ratio of 2. To estimate the total mass $M_\text{tot}$, we use the dynamical mass $M_\text{dyn}$, defined as the total mass within the half-light radius, and estimate $M_\text{tot} \approx 2 M_\text{dyn}$. It has been shown that, for a wide range of density profiles, $M_\text{dyn}$ is well-approximated by~\cite{Wolf:2009tu}
\beq
M_\text{dyn} \simeq 930\,\msun \left(\frac{\sigma_{\text{los}\star}}{\text{km}/\text{s}}\right)^2 \left(\f{R_e}{\text{pc}}\right)
\label{eq:master}
\eeq
where $\sigma_{\text{los}\star}$ is the line-of-sight velocity dispersion of the stars. We select GCs and dwarfs for which the line-of-sight velocity dispersion (or dynamical mass) is available in the catalogs. This results in 62 GCs  and 87 dwarf galaxies. For those dwarf galaxies, the ratio of $M_{\text{tot}\star}/M_{\text{tot}}$ ranges from $4\times 10^{-4}$ to $0.7$ with a median of 0.02.

The resulting scatter plot of $M_\text{dyn}$ vs. $t_0^h$ is shown as the left panel of~\figref{trh_ML}. We use red dots to represent GCs that appear to have core collapsed,\footnote{Observationally, collapsed status is often identified by checking if the density profile of the system cannot be well fitted by a Plummer profile, or any other cored density profile~\cite{2005ApJS..161..304M}. Here, we adopted the status provided by Harris GC catalog~\cite{1996AJ....112.1487H, Harris:2010ut}.} blue stars to represent GCs that have not core collapsed, and black squares to represent the dwarfs. The horizontal band indicates the typical age of the systems, $t_\text{age}=10-13$ Gyr. 
As the figure shows, most of the GCs are around the collapsed phase $t_0^h \simeq t_\text{age}$; this includes all the GCs that have been observed to have core collapsed. On the other hand, the majority of dwarf galaxies have $t_0^h$ much greater than $t_\text{age}$, and thus they show almost no core collapse (manifested as dense stellar cores) or mass segregation. Exceptions include those dwarfs with the smallest stellar masses and stellar half-mass radii. 

Next, we consider a scenario where DM contains two particle species, $\chi$ and $\psi$, whose masses satisfy the hierarchy $m_\chi \ll m_\star \ll m_\psi$. Such a multiple DM scenario is well-motivated for PBHs whose mass function is concentrated around several still-viable mass windows~\cite{Carr:2021bzv}. Such mass functions arise naturally in models such as~\cite{Cai:2018tuh}. We focus on an extreme case in which DM in dwarf galaxies is dominated by the heavier species $\psi$, while GCs are dominated by the lighter species $\chi$. In dwarf galaxies, therefore, the heavier component is $\psi$ rather than stars. Choosing a representative PBH window mass of $m_\psi = 20\,\msun$, the characteristic dynamical evolution timescale for a dwarf galaxy is 
\begin{align}
t_0^h ={}& \frac{(4\pi)^{1/2} \rho_{s\psi}^{1/2} r_{s\psi}^3 }{3 G^{1/2}m_\psi \ln  (2\gamma M_\text{tot}/m_\psi)}\nonumber\\
={}& \frac{4.3\times 10^5\, \text{yr}}{\ln (2\gamma M_\text{tot}/m_\psi)} \left(\frac{20\,\msun}{m_\psi}\right)  \left(\frac{M_{\text{tot}\psi}}{\msun}\right)^{1/2}\left(\frac{r_{s\psi}}{\text{pc}}\right)^{3/2}.
\label{eq:dwarf_heavy}
\end{align}
For GCs, the relevant timescales remain those given in~\eqref{t0_GC} under our assumption.

Once again, we use~\eqref{master} to estimate the dynamical mass and the total mass of a dwarf galaxy, adopting $M_{\text{tot}\psi} \approx M_\text{tot}- M_{\text{tot}\star}$. Because the scale radius of the heavier DM component, $r_{s\psi}$, is not directly observable, we infer it from the concentration–mass–redshift relation.  We first fix the  halo radius, $R_{200c}$, by requiring the averaged density within it approximately equals  200 times the critical density today, $M_{\text{tot}\psi}/(\frac{4}{3}\pi R_{200c}^3) \approx 200 \rho_\text{crit}$. We then use the concentration-mass-redshift relation to determine the concentration $c_{200c}$ for a DM halo of mass $M_{\text{tot}\psi}$ at redshift $z\approx 0$. The scale radius then follows from $r_{s\psi} \approx R_{200c}/c_{200c}$. In practice, we compute $c_{200c}$ with the \texttt{Colossus} package~\cite{Diemer:2017bwl}, adopting the \texttt{ludlow16} model~\cite{Ludlow:2016ifl} and assuming the \texttt{Planck15} cosmology~\cite{Planck:2015fie}. 

We should now summarize the relative contributions of the stellar and DM component(s) to the total mass that underlie the left and right panels of~\figref{trh_ML}, given $M_{\text{tot}\star}/ M_\text{tot} \approx 1$ for the GCs and from $4 \times 10^{-4}$ to $0.7$ with a median of $0.02$ for the dwarfs. In the left panel, the ratio $M_{\text{tot}\chi}/M_\text{tot} \approx (1- M_{\text{tot}\star}/ M_\text{tot}) \ll 1$ for GCs and from 0.3 to 1 with a median of 0.98 for dwarfs. In the right panel, GCs follow
$M_{\text{tot}\psi}/M_\text{tot} \ll M_{\text{tot}\chi}/M_\text{tot} \ll 1$, while dwarfs satisfy $M_{\text{tot}\chi}/M_\text{tot} \ll M_{\text{tot}\psi}/M_\text{tot}$ with $M_{\text{tot}\psi}/M_\text{tot}\approx 1- M_{\text{tot}\star}/M_\text{tot}$ varying from  0.3 to 1 with a median of 0.98.

The right panel of~\figref{trh_ML} shows that
for dwarf galaxies, the characteristic timescale $t_{0}^{h}$ varies almost linearly with the dynamical mass, $M_{\mathrm{dyn}}\approx M_{\text{tot}\psi}$, with a scatter much less than that of the left panel. This behavior follows from $t_0^h \propto M_{\text{tot}\psi}^{1/2}r_{s\psi}^{3/2} \propto M_{\text{tot}\psi}^{1/2} ({M_{\text{tot}\psi}^{1/3}}/{c_{200c}})^{3/2}$, where the concentration changes only modestly across the halo‑mass range considered ($c_{200c}\simeq 14$–$22$).
As a result, core collapse occurs only in dwarf galaxies  with the smallest 
dynamical mass, producing dense dark cores. This is expected as the relaxation time is of $\mathcal{O}(N / \ln (\gamma N))$ times the dynamical time, where $N$ is the number of particles. We have repeated the analysis with several other concentration-mass-redshift models, available in \texttt{Colossus} and applicable for the halo-mass range, and reached the same qualitative conclusion. Although such collapsed dwarf galaxies are expected to be rare in CDM, they have promising observational signatures, including substantial strong lensing effects arising from the ultradense \emph{dark} cores and their cuspy outer density profiles.

\section{Discussion and Conclusion}
\label{sec:conclusion}

In this work, we presented a two-fluid conduction scheme to investigate the dynamical evolution of GCs and dwarf galaxies, focusing on mass segregation and secular core collapse.  The approach treats each component, such as stars and DM, as a separate conducting fluid and solves for their coupled gravothermal evolution. These fluid evolution equations, which we integrated by a semi-implicit finite-difference scheme, can be considered moments of the Boltzmann equation, with the conduction and interaction terms serving as approximations to the collision terms. Such a smooth fluid treatment divided into Lagrangian zones overcomes some of the traditional drawbacks of
N-body simulations, especially in high-density core regions. Numerical benchmarks show that the two-fluid conduction scheme not only reproduces the general features observed in 
N-body results but also provides insight into the late-time collapse behavior of dense systems.

A key result of the study is that the timescales of gravothermal core collapse ($t_c$) and significant mass segregation ($t_s$) largely depend on the fiducial timescale of the heavier component $t_0^h$ (close to the half-mass relaxation time), which is largely determined by its initial properties. By using observations to evaluate such timescales, we conclude that most GCs may experience significant dynamical evolution, whereas for dwarf galaxies, such evolution is much less evident unless they have comparatively small  mass and high compactness at the time of formation.

The degree to which the two fluid profiles evolve depends on the relative scale densities and radii of the lighter and heavier components, as well as the age of the system. By the time core collapse occurs, every system develops an extremely dense core composed of the heavier fluid, surrounded by a quasi-isothermal envelope of the heavier fluid. The lighter fluid expands, but the magnitude of this expansion depends on the initial scale density ratios ($\xi$) and the initial scale radius ratios ($\zeta$). As discussed in~\secref{two-fluid-hierarchy}, for halos with small values of $\xi$ and $\zeta$, their lighter fluid's Lagrangian radii are increased by roughly a factor of a few (see \figref{pdmB}), while their central densities are reduced  by 2--4 orders of magnitude (see \figref{pdm}). In contrast, for halos with large $\xi$ and $\zeta$, the lighter fluids undergo only modest expansions, and the corresponding central densities drop by no more than a factor of a few.

Based on the simulations in~\secref{two-fluid-hierarchy} and discussions in~\secref{observation}, we conclude that, although the dynamical heating could significantly lower the central density of the lighter DM in a GC, the process is unlikely to significantly deplete the initial DM distribution at larger radii and cause the apparent absence of DM in GCs, even though such heating is more important in GCs than in typical UCDs. However, this conclusion will have to be completely re-examined when binary star formation and Galactic tidal interactions are incorporated. The former will increase the heating rate and can reverse the core collapse of the heavier species,  while the latter will make it easier for the lighter DM to escape the cluster. Either or both of these effects might reverse our conclusion.

Looking ahead, our formalism and code can be adapted to more realistic stellar–DM configurations with low computational cost. It could be extended to include tidal interactions or additional components such as binaries or massive central black holes, offering a way to assess how these elements influence the evolution of stellar–DM systems. Furthermore, although we focus on systems with a single stellar and a  CDM (PBH) population in this work, the framework can be applied to SIDM halos or halos with multiple stellar populations or PBH populations. This versatility may help clarify the nature of DM and the physics of dense astrophysical structures. We hope to explore some of these effects in future investigations.

\subsection*{Acknowledgements}
We thank Anirudh Chiti, Liang Dai, Ariane Dekker, Alex Drlica-Wagner, Moritz Fischer, Alexander Ji, Andrey Kravtsov, Burcin Mutlu-Pakdil,   Ethan Nadler, Kaixiang Wang, and Hai-Bo Yu for useful discussions. The simulations were partly performed at the University of Chicago's Research Computing Center. YZ acknowledges the Aspen Center for Physics, which is supported by NSF Grant No. PHY-1607611, and the Kavli Institute for Theoretical Physics, which is supported by NSF Grant No. PHY-2309135, for their hospitality during the completion of this study. YZ is supported by the GRF Grants No. 11302824 and No. 11310925 from the Research Grants Council of Hong Kong SAR and the Grant No. 9610645 from the City University of Hong Kong. SLS acknowledges support from NSF Grants No. PHY 2006066 and No. PHY 2308242 to the University of Illinois at Urbana-Champaign.

\bibliography{simulation}
\bibliographystyle{h-physrev.bst}

\appendix
\section{Algorithm for the conduction/interaction step}
\label{sec:CFL}
Our conduction fluid simulation involves solving partial differential equations, such as the equation of conduction/interaction, which includes a first-order derivative in time and a second-order derivative in space. For the solution to be stable, our algorithm must satisfy the CFL condition. The condition can typically  be achieved by carefully selecting the spatial and temporal intervals in numerically solving the partial differential equation.  Ideally, by carefully designing the algorithm, the CFL condition can be satisfied for arbitrary  choice of spatial and temporal intervals, i.e., it is 
unconditionally stable. Below, we will introduce a semi-implicit algorithm to solve the conduction/interaction equation and prove that it is unconditionally stable whenever $m_l \gg m_h$. 

\subsection{A semi-implicit algorithm to solve the conduction/interaction equation}

Let us first focus on the conduction/interaction of the heavier fluid.  Substituting~\eqref{starfid7} into~\eqref{starfid6} yields
\begin{align}
 \frac{D\t u_h}{D\t t} ={}&  \f{c'_2}{\t \rho_h \t r^2}\frac{\partial}{ \partial \t r} \left( \t \rho_h \t r^2 \frac{\partial \sqrt{\t u_h}}{\partial \t r}\right) -\f{c_1 \t \rho_l \left(\t u_h  - \frac{m_l}{m_h} \t u_l\right)}{(\t u_h+ \t u_l)^{3/2}} \nonumber\\
={}& \f{c'_2}{\t r^2}\left[\left(1+\f{\partial \ln \t \rho_h}{\partial \ln \hat r}\right)\f{\partial \sqrt{\t u_h}}{\partial \ln \t r}+\f{\partial^2 \sqrt{\t u_h}}{(\partial \ln \t r)^2}\right] \nonumber\\
&-\f{c_1 \t \rho_l \left(\t u_h  - \frac{m_l}{m_h} \t u_l\right)}{(\t u_h+ \t u_l)^{3/2}},
\label{eq:conduction_simp}
\end{align}
where $c'_2\equiv 2 c_2 {\ln \Lambda_h}/{\ln \Lambda_{hl}}$. If we use an explicit algorithm to solve the partial differential equation~\eqref{conduction_simp}, the  stability is not always guaranteed for arbitrary choice of $\Delta \hat t$ and $\Delta \ln \hat r$. 

To improve the situation, we instead use a semi-implicit algorithm to solve the conduction/interaction equation. Specifically, we evaluate some of the $\hat u$ variables on the right-hand side of~\eqref{conduction_simp} implicitly, i.e.,
\begin{align}
&\frac{\hat{u}_{h,j}^{n+1} - \hat{u}_{h,j}^n}{\Delta \hat{t}} 
= \f{c'_2}{\t r^2}\left[\left(1+\f{\Delta \ln \t \rho^n_h}{\Delta \ln \hat r}\right)\left.\f{\partial \sqrt{\t u_h}}{\partial \ln \t r}\right|_{j, n+1}\right. \nonumber\\
&\left.+\left.\f{\partial^2 \sqrt{\t u_h}}{(\partial \ln \t r)^2}\right|_{j, n+1}\right]-\f{c_1 \t \rho^n_{l,j} \left(\t u^{n+1}_{h,j}  - \frac{m_l}{m_h} \t u^{n+1}_{l,j}\right)}{(\t u^n_{h,j}+ \t u^n_{l,j})^{3/2}},
\label{eq:conduction_simp}
\end{align}
where the subscript $j$ is the zone index, and the superscript $n$ is the time step index (should not be confused with power). To get the value of $\left(\sqrt{\hat u_h}\right)^{n+1}_j$, we apply an Taylor expansion around $\sqrt{\hat u}$ at $n$-th step,
\begin{align}
\left(\sqrt{\hat u_h}\right)^{n+1}_j\simeq {}& \left(\sqrt{\hat u_h}\right)^{n}_j + \left.\frac{\partial \sqrt{\hat u_h}}{\partial \hat u_h}\right|_{j,n} (\hat{u}_{h,j}^{n+1} - \hat{u}_{h,j}^n) \nonumber\\
  ={}& \sqrt{\hat{u}_{h,j}^n}+ \frac{1}{2\sqrt{\hat{u}^n_{h,j}}}(\hat{u}_{h,j}^{n+1} -\hat{u}_{h,j}^n) 
\label{eq:changevariable}
\end{align}
Therefore, we can approximate
\begin{align}
\left.\frac{\partial \sqrt{\hat u_h}}{\partial \ln \hat{r}}\right|_{j,n+1} \approx{}& \frac{\hat{z}_{h, j+1/2}^{n+1}-\hat{z}_{h, j-1/2}^{n+1}}{\Delta \ln \hat{r}},\\
\left.\frac{\partial^2 \sqrt{\hat u_h}}{(\partial \ln \hat{r})^2}\right|_{j,n+1} \approx{}&
\frac{\hat{z}_{h, j+1}^{n+1}- 2 \hat{z}_{h, j}^{n+1} + \hat{z}_{h, j-1}^{n+1}}{(\Delta \ln \hat{r})^2},
\end{align}
where $\hat{z}_{h, j}^{n+1}  \equiv \sqrt{\hat{u}_{h,j}^n}+ \frac{1}{2\sqrt{\hat{u}^n_{h,j}}}(\hat{u}_{h,j}^{n+1} -\hat{u}_{h,j}^n)$.

The conduction/interaction equation for the lighter fluid is given by
\begin{align}
\f{D\t u_l}{D\t t} ={}&\frac{m_l}{m_h}  \frac{c''_2}{\hat \rho_l \hat r^2}\frac{\partial }{\partial \t r} \left(\t \rho_l \t r^2  \f{\partial \sqrt{\t u_l}}{\partial \t r}\right) +   \frac{c_1 \t \rho_h  \left(\t u_h- \frac{m_l}{m_h} \t u_l\right)}{(\t u_h +\t u_l)^{3/2}}, \nonumber\\
={}& \frac{m_l}{m_h} \frac{c''_2}{\hat r^2} \left[\left(1+\f{\partial \ln \t \rho_l}{\partial \ln \hat r}\right)\f{\partial \sqrt{\t u_l}}{\partial \ln \t r}+\f{\partial^2 \sqrt{\t u_l}}{(\partial \ln \t r)^2}\right] \nonumber\\
&+\f{c_1 \t \rho_h \left(\t u_h  - \frac{m_l}{m_h} \t u_l\right)}{(\t u_h+ \t u_l)^{3/2}},
\label{eq:conduction_l}
\end{align}
where $c''_2 \equiv 2 c_2 {\ln \Lambda_l}/{\ln \Lambda_{hl}}$. The discretization can be done in a similar manner as that of the heavier fluid.

In the limit $m_l \ll m_h$,~\eqsref{conduction_simp}{conduction_l} can be respectively discretized into
\begin{align}
&\frac{\hat{u}_{h,j}^{n+1} - \hat{u}_{h,j}^n}{\Delta \hat{t}} = \frac{c'_2}{\hat{r}^2} \left[\left(1+\frac{\Delta \ln \hat{\rho}^n_h}{\Delta \ln \hat{r}}\right)\frac{\hat{z}_{h, j+1/2}^{n+1} - \hat{z}_{h, j-1/2}^{n+1}}{\Delta \ln \hat{r}}\right.\nonumber\\
&\left. + \frac{\hat{z}_{h, j+1}^{n+1} - 2 \hat{z}_{h,j}^{n+1} + \hat{z}_{h, j-1}^{n+1}}{(\Delta \ln \hat{r})^2} \right] -\f{c_1 \hat \rho^n_{l,j} \hat u^{n+1}_{h,j} }{(\hat u_{l,j}^n+ \hat u^n_{h,j})^{3/2}}
\label{eq:simple_h}
\end{align}
\begin{equation}
\frac{\hat{u}_{l,j}^{n+1} - \hat{u}_{l,j}^n}{\Delta \hat{t}} = \f{c_1 \hat \rho^n_{h,j}  \hat u^{n+1}_{h,j} }{(\hat u_{l,j}^n+ \hat u^n_{h,j})^{3/2}}.
\label{eq:simple_l}
\end{equation}

\subsection{Checking the CFL condition}
In the limit $m_l \ll m_h$, only the heavier fluid contains the conductivity term, see~\eqref{simple_h}, which involves spatial derivative requires checking the CFL condition. We start by imposing the standard modal ansatz
\beq
\hat u_{h,j}^n = (\xi_h)^n e^{\text{i}k j \Delta \ln \hat r},
\label{eq:ansatz}
\eeq
and substitute it to~\eqref{simple_h}. If the magnitudes of  $\hat u_h$'s are always bounded by unity, i.e., $|\xi_h|<1$, for all the wavenumbers $k$, then the CFL condition for stability is satisfied. 

Under~\eqref{ansatz}, the auxiliary variable is given by
\begin{align}
 z^{n+1}_{h,j} \equiv{}& \sqrt{\hat{u}_{h,j}^n}+ \frac{1}{2\sqrt{\hat{u}^n_{h,j}}}(\hat{u}_{h,j}^{n+1} -\hat{u}_{h,j}^n) \nonumber\\
 ={}&\sqrt{\hat{u}_{h,j}^n} \frac{\xi_h+1}{2}
 \label{eq:ansatz2}
\end{align}
Substituting~\eqsref{ansatz}{ansatz2} into~\eqref{simple_h} leads to
\begin{align}
\xi_h -1 & =\f{c'_2 (\xi_h+1) \Delta \hat t}{\hat r^2 \sqrt{\hat u_{h,j}^n}} \left[\left(1+\f{\Delta \ln \hat \rho^n_{h}}{\Delta \ln \hat r}\right)\f{\text{i} \sin \frac{k \Delta \ln \hat r}{4}}{\Delta \ln \hat r}  \right.\nonumber\\
&\left.-\f{2 \sin^2 \frac{k\Delta \ln \hat r}{4}}{(\Delta \ln \hat r)^2}\right] - \frac{c_1 \hat \rho^n_{l,j}  \Delta \hat t }{(\hat u_{h,j}^n + \hat u_{l,j}^n)^{3/2} } \xi_h 
\label{eq:twofluidCFL3}
\end{align}
Let
\begin{align}
A \equiv{}& \f{c'_2  \Delta \hat t}{\hat r^2 \sqrt{\hat u_{h,j}^n}} \left(1+\f{\Delta \ln \hat \rho^n_{h}}{\Delta \ln \hat r}\right)\f{\sin \frac{k \Delta \ln \hat r}{4}}{\Delta \ln \hat r}, \nonumber \\
B \equiv{}& \f{2 c'_2  \Delta \hat t}{\hat r^2 \sqrt{\hat u_{h,j}^n}} \f{\sin^2 \frac{k\Delta \ln \hat r}{4}}{(\Delta \ln \hat r)^2}, \nonumber\\
C \equiv{}& \frac{c_1 \hat \rho^n_{l,j}  \Delta \hat t }{(\hat u_{h,j}^n + \hat u_{l,j}^n)^{3/2} },
\end{align}
where $B$ and $C$ are always greater than 0 while the sign of $A$ is undetermined. 
\eqref{twofluidCFL3} can be re-expressed as 
\beq
\xi_h -1 = (\xi_h +1) (A\text{i} - B)- C \xi_h
\eeq
The magnitude square of the resulting $\xi_h$ is given by
\beq
|\xi_h|^2 = \frac{A^2 + (B-1)^2}{A^2 + (B+C+1)^2},
\eeq
Given $(B-1)^2 - (B+C+1)^2 = (2B+C) (-2-C)< 0$, we have
\beq
|\xi_h| < 1
\eeq 
for an arbitrary choice of $\Delta \ln \hat r$ and $\Delta \hat t$. Therefore, the algorithm is  unconditionally stable.

Beyond the limit of $m_l \ll m_h$, both the heavier and lighter fluids contain the conductivity term, and we need to check the CFL condition for the coupled system. The discussion can be conducted in a similar manner as that of the $m_l \ll m_h$ limit, albeit we require the magnitudes of both $\hat u_l$ and $\hat u_h$ to be bounded by unity. In such a scenario, our algorithm is no longer unconditionally stable, and one needs to set a limit on the time step for the adopted grid spacing.

\section{Translation of the fiducial units}
\label{sec:trans}

Ref.~\cite{Shapiro:2018vju} uses a different set of fiducial quantities. Here, we translate its fiducial quantities to the ones used in the main text. For the Plummer profile~\eqref{plummer} with scale radius $r_{sh}$ and the scale density $\rho_{sh}$, the fiducial quantities used in~\cite{Shapiro:2018vju} (with tildes) can be expressed as
\begin{align}
   \tilde R_{0} \equiv {}& 2^{-1/2} r_{sh}, \\
   \tilde M_{0} \equiv {}& M_\text{tot} = \f{4\pi \rho_{sh} r_{sh}^3}{3} = \frac{M_0}{3},\\
     \tilde \rho_{0} ={}& \frac{\tilde M_{0}}{4\pi \tilde R_{0}^3} = \f{2^{3/2}}{3} \rho_{sh},\\
   \tilde \sigma_{0} ={}& \left(\frac{G \tilde M_0}{\tilde R_0} \right)^{1/2}= \f{2^{1/4}}{\sqrt{3}} r_{sh} (4\pi G \rho_{sh})^{1/2} =  \f{2^{1/4}}{\sqrt{3}} \sigma_0,\\
   \tilde t_{0} = {}& \f{1}{6 b} \f{3^{3/2} \tilde \sigma_{0}^3}{4\pi G^2 \alpha \tilde \rho_{0} m_h \ln \Lambda_h}  \nonumber\\
  ={}&\f{3}{2^{7/4} \alpha b} \frac{\sigma_0^3}{12 \pi G^2 \rho_0 m_h \ln \Lambda_h }= 1.629\, t_0, 
  \label{eq:trans}
\end{align}
where we explicitly put back the gravitational constant $G$. Ref.~\cite{Shapiro:2018vju} reports the time of the snapshot of the density profile in the caption of Fig.~1. The central density of the last snapshot exceeds $10^6 \tilde \rho_0$, which is well into core collapse. The corresponding time is given by $t_\text{last}/\tilde t_0 = 3.402$. According to~\eqref{trans}, the time can be translated into our fiducial time convention by
\beq
\hat t_\text{last} = \frac{t_\text{last}}{t_0} = \frac{t_\text{last}}{\tilde t_0} \frac{\tilde t_0}{t_0} = 5.542.
\eeq
Therefore, we expect the collapse time of the one-fluid simulation of~\cite{Shapiro:2018vju}, defined where $\rho_c \rightarrow \infty$, to be only slightly longer than $\hat t_\text{last}$. 
Our result using the two-fluid scheme in the one-fluid limit (see~\tabref{tab1}) is in reasonable agreement with this expectation. 

\section{Comparing the two-fluid conduction and the Fokker-Planck simulations}
\label{sec:FP}

\begin{figure*}[t]
   \centering
   \includegraphics[width=0.96\textwidth]{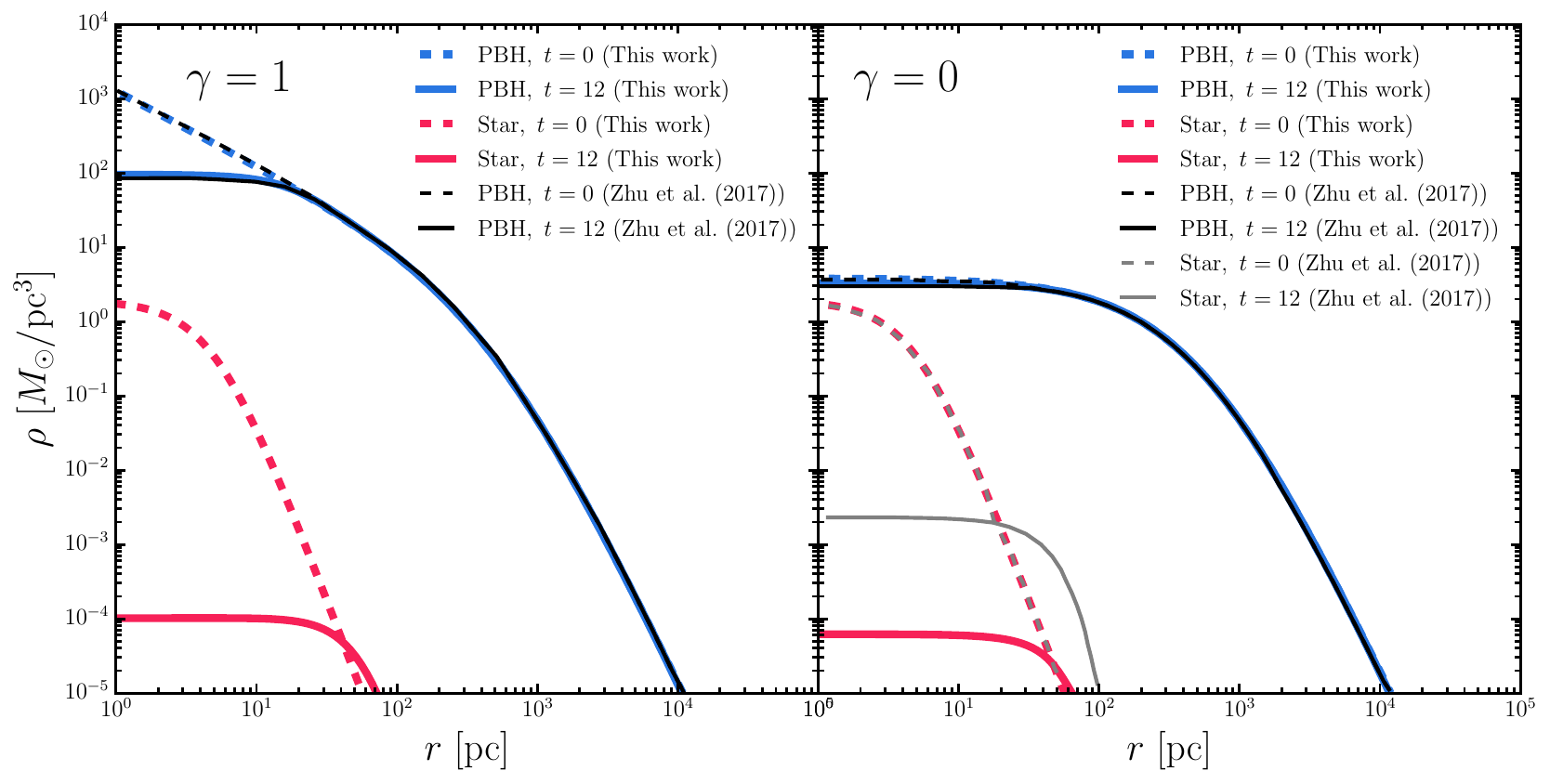} 
   \caption{(Left) Snapshots of the density profiles for a PBH–stellar system. The dashed lines show 
the initial PBH and stellar profiles, given by the Dehnen profile with $\gamma=1$ and 
the Plummer profile, respectively. The solid lines show the density profile at 
$t=12\,\text{Gyr}$. Blue ($t=0$) and red ($t=12$) lines denote the fluid model results, while black lines ($t=0$)
indicate those from the FP model. (Right) The same setup as in the left panel, except that the initial PBH profile is given 
by the Dehnen profile with $\gamma=0$. The blue ($t=0$) and red  ($t=12$) lines represent the fluid model results, and the black  ($t=0$) and gray  ($t=12$) lines represent the FP model results.
}
   \label{fig:FP}
\end{figure*}

Ref.~\cite{Zhu:2017plg} studied the evolution of ultra-faint dwarf galaxies composed of stars and PBHs using FP simulations. Here, we re-run the evolution of such systems with our fluid model and compare the resulting profiles with those from the FP model. According to~\cite{Zhu:2017plg}, initially the stars are described by a Plummer profile~\eqref{plummer}, while the PBHs follow a Dehnen profile~\cite{Dehnen:1993uh}:
\begin{equation}
\rho(r) = \frac{(3-\gamma)\,M_{\text{tot}\psi}}{4\pi\,r_{s\psi}^3}
\left(\frac{r}{r_{s\psi}}\right)^{\gamma}
\left(1 + \frac{r}{r_{s\psi}}\right)^{\gamma - 4},
\end{equation}
where the inner slope parameter $\gamma$ is fixed to 0 or 1 for the relaxed or non-relaxed PBH populations, respectively. To set up the simulation, we follow the choices in Ref.~\cite{Zhu:2017plg}:
\begin{align}
m_\star ={}& 1\,\msun,\quad
M_{\text{tot}\star} = 10^3\,\msun,\quad
r_{s\star} = 5\,\text{pc}, \nonumber\\
m_\psi ={}& 30\,\msun,\quad
M_{\text{tot}\psi} = 2\times 10^9\,\msun,\quad
r_{s\psi} = 500\,\text{pc}.\nonumber
\end{align}
With these parameters, the PBHs behave as the heavier fluid and the stars as the lighter one. In our fluid simulation, we again set the conduction coefficients to $\beta_{\star} = \beta_\psi = 1$.

The resulting evolution of the density profiles is shown in~\figref{FP}, starting from either a cuspy PBH profile combined with a cored stellar profile (left panel) or both a cored PBH and cored stellar profile (right panel). In both cases, the stellar fluid expands while the PBH fluid either smooths its inner cusp (left panel) or undergoes a very mild expansion (right panel). These features agree well with the FP simulation results,  which are digitized from Figs.~1 and~2 of~\cite{Zhu:2017plg} and plotted as the black and gray lines in~\figref{FP}. Note that~\cite{Zhu:2017plg} does not provide the snapshot for the stellar evolution for the $\gamma=1$ scenario. The only exception is the stellar evolution at $t=12$ Gyr, where the FP model exhibits an evolution leading to a milder expansion and denser core than the fluid model (gray solid vs. red solid) in the right panel.  Such discrepancy may be due to uncertainties in the conduction coefficients, which could be calibrated in principle; we leave the calibration for future work.

\end{document}